\documentclass[11pt,oneside]{article}
\usepackage{setspace,graphicx,amssymb,amsmath,latexsym,amsfonts,amscd,amsthm,multirow,ctable,mathdots,caption,array,diagbox,mathtools}
\usepackage{chet}
\usepackage{tabularx,cite,mathrsfs}
\usepackage{authblk}
\usepackage{color}

\usepackage{fullpage}
\usepackage{stmaryrd}
\usepackage{rotating}

\usepackage{hyperref}

\newcommand{\be}{\begin{eqnarray}}
\newcommand{\ee}{\end{eqnarray}}
\newcommand{\bea}{\begin{align}}
\newcommand{\eea}{\end{align}}
\newcommand{\nn}{\nonumber}
\newcommand{\bb}{\mathbb}
\newcommand{\mc}{\mathcal}
\def\({\left(}
\def\){\right)}

\renewcommand{\th}{\theta}

\newcommand{\Tr}{{\rm Tr \,}}

\newcommand{\es}[2] {\begin{equation} \label{#1} \begin{split} #2 \end{split} \end{equation}}

\newcommand{\tr}{\mathrm{Tr}}

\begin{document}

\setstretch{1.4}

\title{\vspace{-65pt}
\vspace{20pt}
    \textsc{\huge{
    Universal Bounds on Charged States in 2d CFT and 3d Gravity}
    \\
    }}

\author[a]{Nathan Benjamin}
\author[a]{Ethan Dyer}
\author[b]{A. Liam Fitzpatrick}
\author[a]{Shamit Kachru}
\affil[a]{Stanford Institute for Theoretical Physics, 
Via Pueblo, Stanford, CA 94305, USA}
\affil[b]{Boston University Physics Department, 
Commonwealth Avenue, Boston, MA 02215, USA}

\date{}

\vspace{-1em}

\abstract{We derive an explicit bound on the dimension of the lightest charged state in two dimensional conformal field theories with a global abelian symmetry. We find that the bound scales with $c$ and provide examples that parametrically saturate this bound. We also prove that any such theory must contain a state with charge-to-mass ratio above a minimal lower bound.  We comment on the implications for charged states in three dimensional theories of gravity.
}

\maketitle

\clearpage

\tableofcontents

\section{Introduction}

At low energies, theories of quantum gravity are straightforward to formulate quantitatively as effective field theories.  However, the universal coupling of gravity implies that, unlike gauge theories, at high energies gravity has no decoupling limit where it can be separated from the matter content of the theory.  This might lead one to expect that precise statements about the spectrum or dynamics of gravity at high energies would necessarily be contingent on some knowledge of the low-energy spectrum.  Conversely, insisting on various criteria for the behavior of the ultraviolet description ought to impose non-trivial constraints on the behavior of the theory at low-energies.  

Holographic approaches allow one to make this intuition precise and to extract quantitative predictions by turning a poorly-defined question, that of how to formulate the space of UV-complete theories of gravity, into sharp questions about observables at the boundary of space-time.  In flat space,  the corresponding observable is the S-matrix, and the implications of analyticity and other axioms of the S-matrix provide a path to constraining the dynamics of the theory \cite{Adams:2006sv}.  Turning on a small negative cosmological constant, the theory apparently remains nearly unchanged locally; yet the global Anti de Sitter space-time structure is radically different, and the boundary observables now comprise the full dynamics of a Conformal Field Theory \cite{GKP,Witten,Maldacena:1997re}.  In fact, the structure of CFTs is sufficiently rigid that it is possible to derive universal constraints on the dynamics and spectrum of all theories of gravity in AdS at low and high energies \cite{Fitzpatrick:2014vua,Fitzpatrick:2010zm,Komargodski:2012ek,Fitzpatrick:2012yx,Alday:2007mf,Hellerman,Hartman:2014oaa,Benjamin:2015vkc,Benjamin:2015hsa}.  Such constraints clearly can also be applied to the full space of CFTs, which is a central area of study in its own right. 

The most powerful such methods are found in the case of the correspondence between three-dimensional gravity in Anti de Sitter space and two-dimensional Conformal Field Theories, i.e. AdS$_3$/CFT$_2$.  All graviton degrees of freedom are purely boundary excitations whose dynamics are completely fixed by the infinite conformal symmetry in two dimensions.  Moreover, modular invariance of the theory at finite temperature relates the spectrum at high energies and the spectrum at low energies. This makes it possible to draw sharp conclusions about the high-energy dynamics as a function of the assumptions about the low-energy spectrum. One can even derive  properties that are common to all gravitational theories, i.e. that make only very basic assumptions such as unitarity.  
 
A remarkable result along these lines was derived in \cite{Hellerman} and systematically improved upon numerically in \cite{Friedan:2013cba,Qualls:2015bta,Qualls:2014oea,Qualls:2013eha}, where a rigorous and universal upper bound was found on the mass $m_L$ of the lightest bulk degree of freedom in AdS$_3$.  Roughly, this bound is $m_L \lesssim \frac{1}{4 G_{\rm N}}$, where $G_{\rm N}$ is Newton's constant.    When the mass of a state is greater than the threshold $m_{\rm BH} \ge \frac{1}{8G_{\rm N}}$ for black holes in AdS$_3$ \cite{BTZ}, classical gravity predicts that it should collapse and form a horizon, and we can suggestively call it a ``black hole'' state.  A priori, there is no guarantee that any specific solution to Einstein's equations is actually a physical state in all theories of quantum gravity; in fact, most are not, since the spectrum of states in a CFT is usually discrete whereas the spectrum of solutions to general relativity (GR) is continuous.  The results of \cite{Hellerman,Friedan:2013cba} can therefore very roughly be summarized as the statement that the spectrum of states in theories of gravity must, at a bare minimum, contain black holes near threshold.  Moreover, the result applies to {\it all} CFTs, even those whose bulk duals in AdS$_3$ may not be well-described by GR at high energies, or where the curvature of AdS$_3$ is order ${\cal O}(1)$ in units of the Planck scale.   

Our main goal in this paper will be to extend this technique to the case where the bulk theory has a gauge field in addition to gravity.  This is dual to the assumption that there is conserved vector current $J$ in the boundary CFT.  Now, one can ask not only about the spectrum of energies of states, but also about their charges.  A very simple but elusive question is whether there must be charged states in the theory at low energies, or if instead they can be made arbitrarily heavy and therefore effectively decoupled from any description at fixed finite energy.  Charged black holes would seem to exist as classical solutions in GR with a gauge field, but it is not obvious that a theory with only neutral states is actually inconsistent, since in such a theory charged black holes can never be produced.   Various arguments have been made that light charged states must be present below some bound in mass\cite{ArkaniHamed:2006dz,Banks:2010zn,Harlow:2015lma}, but so far a rigorous proof is lacking.  A key result of this paper will be to prove such an upper bound.

This result has a clear connection to the Weak Gravity Conjecture (WGC) of \cite{ArkaniHamed:2006dz}, which exists in multiple forms but in any case is an upper bound on $\frac{m}{Q m_{\rm pl}}$ for the mass $m$ and charge $Q$ of some state in the theory.\footnote{One might expect that the WGC should be qualitatively modified in AdS$_3$ due to peculiarities of three dimensions.  In  particular, in AdS$_3$, the relation between mass and charge of extremal BHs is qualitatively different, $M \sim Q^2$.  Furthermore, boundary currents in 2d are dual to Chern-Simons gauge fields in AdS$_3$, which have no bulk degrees of freedom. On the other hand, if the WGC is sufficiently robust under compactification of extra dimensions, one might expect these peculiarities to be irrelevant. See \cite{Heidenreich:2015nta} for discussions of which versions of the WGC are robust to compactification of dimensions, and \cite{Nakayama:2015hga} for discussions of how WGC might be modified in AdS/CFT contexts. In any case, our bounds are a rigorous consequence of modular invariance and thus provide an independent approach to studying the WGC in AdS$_3$.}  At large central charge $c$, for non-chiral (which have, e.g., $c=\bar{c}$) CFTs our upper bound on the weight $\Delta$ of the lightest charged state in the theory asymptotes to
\be
\Delta - \Delta_{\rm vacuum} <   \frac{c}{6}+ \frac{3}{2\pi} + {\cal O}\left( \frac{1}{c} \right), 
\label{eq:ourgap}
\ee
 where $\Delta_{\rm vacuum} = -\frac{c}{12}$ is the weight of the vacuum.   In terms of AdS quantities, this translates to $m \lesssim \frac{1}{4 G_{\rm N}}$ for the lightest charged state.\footnote{
Recall that in GR, the CFT central charge satisfies $c = \frac{3 \ell_{\rm AdS}}{2 G_{\rm N}}$ \cite{Brown:1986nw}, and for a bulk scalar the weight satisfies $(\Delta-\Delta_{\rm vacuum})(\Delta-\Delta_{\rm vacuum}-2) = m^2 \ell_{\rm AdS}^2$.}
 The bound is  numerically determined from the modular bootstrap and can likely be improved with better numerics.   We will also find a bound on $\frac{m G_{\rm N}}{Q}$ when $Q$ is normalized so that the level $k$ of the current $J$ is $1$.\footnote{Equivalently, one can keep $k$ explicit and replace $Q$ with $Q/\sqrt{k}$.  From the CFT point of view, where we do not assume any {\it a priori} knowledge of the charges that arise in the theory, the central charge $k$ shows up only in the two-point function of the current, $J(x) J(0) \sim \frac{k}{x^2}$ and is completely removed by canonically normalizing the currents $J \rightarrow J'= J/\sqrt{k}$ and charges $Q \rightarrow Q'=Q/\sqrt{k}$.   So the actual value of $k$ appears to be invisible to the CFT data we are using, and we can set $k=1$ without loss of generality.  We thank Dan Harlow for encouraging us to emphasize this.  }  At large $c$, it implies that there exists a state in the theory with charge to mass ratio satisfying,
 \be
 \frac{Q}{m G_{\rm N}} > \frac{1}{4 \sqrt{\pi}}, \qquad (c \gg 1)\,.
 \ee
This bound could also likely be improved with further effort.  One might expect or hope for a better bound,
but our main point is that it is parametrically ${\cal O}(1)$ and is a rigorous proof that there must exist some state in the theory with $ \frac{m G_{\rm N}}{4 \sqrt{\pi}}< Q$.

It is notable that the upper bound (\ref{eq:ourgap}) is at $\sim \frac{c}{6}$ and not $\sim \frac{c}{12}$, as one might expect from the classical threshold for black holes in AdS$_3$.  An analogous mismatch arises in the bounds found in \cite{Hellerman,Friedan:2013cba}, and it is unclear if this is a short-coming of the methods (for instance, only the subgroup $\tau \rightarrow -\frac{1}{\tau}$ of modular transformations is actually used) or if there are physical theories that saturate the weaker bound.  Partly motivated by this, we also consider stronger bounds on the gap to the lightest charged state that can be obtained in the case of $\mc{N}=(1,1)$ supersymmetric theories with a $U(1)$ current.  Here, we can consider a holomorphic quantity called the elliptic genus, and indeed we find the improved bound on the weight of the lightest charged state,
\be
\Delta-\Delta_{\rm vacuum} \le \frac{c}{12} +1 \qquad \textrm{(supersymmetric})\,.
\ee 

One might also wonder how much more the gap to charged states can be lowered in principle by any methods, not necessarily those used here.  In particular, one might hope to prove on general grounds that charged states should enter parametrically below $M_{\rm pl}$.  While this may indeed prove true after restricting to certain classes of theories,\footnote{See e.g. \cite{Harlow:2015lma,Cheung:2014ega} for recent interesting arguments along these lines.  In particular, all examples we present have a coupling for the gauge field that is ${\cal O}(1)$ at the Planck scale, whereas \cite{Harlow:2015lma} argues for a stronger bound only when this gauge coupling is small.} we examine a few counter-examples that demonstrate it cannot be true in complete generality.

The outline of the paper is as follows.  In section \ref{sec:modtrans}, we discuss the transformation of the partition function in the presence of a chemical potential, and the corresponding characters.  In section \ref{sec:bounds}, we derive our bounds on the charged spectrum. In section \ref{sec:largegapex}, we present specific models to demonstrate that our bounds are close to saturating the optimal bounds that are possible at large $c$ without making additional assumptions.  In section \ref{sec:disc}, we discuss potential future directions.

\section{Modular Transformations with Currents}
\label{sec:modtrans}

In any CFT with a conserved current $J$, one can consider the partition function graded by the charge of the current:
\be
Z(\tau,z) &\equiv& \textrm{tr} \left( q^{L_0-c/24} \bar{q}^{\bar{L}_0-\bar{c}/24} y^{J_0}\right)\,,
\label{eq:flavorflav}
\ee
where $q=e^{2 \pi i \tau}$ and $y=e^{2\pi i z}$.  The starting point of our analysis is that under modular transformations,
\be
\tau \rightarrow \tau' = \frac{a \tau + b}{c \tau +d}, \qquad z \rightarrow z' = \frac{z}{c\tau + d},
\label{eq:CoordModTrans}
\ee
the partition function transforms in a universal way:\footnote{Neither $c$ nor $k$ in (\ref{eq:PFModTrans}) are related to the central charge of the theory.  The variable $c$ is from the transformation in (\ref{eq:CoordModTrans}) and $k$ is the level of the current algebra.}
\be
Z(\tau', z') &=& e^{\pi i k \left( \frac{cz^2}{c\tau+d}- \frac{c\bar{z}^2}{c\bar{\tau}+d} \right) } Z(\tau,z).
\label{eq:PFModTrans}
\ee
We will discuss the argument for this transformation below, and work through an illustrative example. 
 
\subsection{Derivation of Transformation}
\label{sec:transderiv}
Most of our analysis in this paper is based on the transformation property of the flavored partition function under a modular transformation, (\ref{eq:PFModTrans}). As we explain in detail in Appendix \ref{app:transformationderivation}, this transformation property is independent of the particular theory, and only depends on the universal structure of $U(1)$ current algebra.\footnote{The explicit form of the partition function is of course theory dependent. It is only the transformation property which is universal. The transformation applies as well to the non-compact abelian group $\mathbb{R}$; in fact, since our analysis makes no assumption about the representations that arise in the theory, it does not distinguish between the cases $U(1)$ and $\mathbb{R}$. For conciseness, however, we will simply refer to the abelian group as ``$U(1)$''.} It is also possible to derive this transformation directly using algebraic properties of the modes $\oint_\gamma dz J(z)$ of the current $J(z)$ on different cycles $\gamma$ of the torus, as shown in \cite{Dijkgraaf:1987vp,Verlinde:1986kw}.\footnote{We thank Herman Verlinde for bringing this argument to our attention.  Their argument is in several ways more satisfying and elegant, and more suggestive of how an argument might be generalized beyond the partition function.}  
Whichever method one prefers, once one knows that the transformation property is theory-independent it becomes sufficient to derive it in a particularly simple theory.  

Let us review this explicitly in the case of  the free boson. For a free boson on a circle of radius $R$, the primary states under the $U(1)$ current algebra are labeled by two integers, $|m,n\rangle$, for momentum and winding. These states satisfy,
\es{states}{
j_{0}|m,n\rangle\,=\,\underbrace{\frac{m}{2R}+nR}_{p_{L}}|m,n\rangle\,,& \ \ \ \bar{j}_{0}|m,n\rangle\,=\,\underbrace{\frac{m}{2R}-nR}_{p_{R}}|m,n\rangle\\
L_{0}|m,n\rangle\,=\,\frac{p_{L}^{2}}{2}|m,n\rangle\,,& \ \ \ \bar{L}_{0}|m,n\rangle\,=\,\frac{p_{R}^{2}}{2}|m,n\rangle\,.
} 

The flavored partition function is then given by,
\es{zbos}{
Z_{\rm bos}(\tau,z)&=\frac{1}{|\eta(\tau)|^{2}}\sum_{m,n\in\mathbb{Z}}q^{\frac{p_{L}^{2}}{2}}\bar{q}^{\frac{p_{R}^{2}}{2}}y^{p_{L}}\bar{y}^{p_{R}}\,.
}
This partition function is invariant under the transformation $\tau\rightarrow\tau+1$. Under the S transformation, $\tau\rightarrow-1/\tau$, the transformation of the partition function can be easily computed by applying the Poisson resummation formula,
\es{poisres}{
\sum_{\ell}e^{-\pi a \ell^{2}+b\ell}=\frac{1}{\sqrt{a}}\sum_{k}e^{\frac{\pi}{a}\left(k+\frac{b}{2\pi i}\right)^{2}}\,,
}
to both the $m$ and $n$ sums. Combining this with the modular transformation property of the $\eta$ function, $\eta(-1/\tau)=\sqrt{i\tau}\eta(\tau)$. We have
\es{transres}{
Z_{\rm bos}(\tau^{\prime},z^{\prime})&=e^{\pi i \left( \frac{z^2}{\tau}- \frac{\bar{z}^2}{\bar{\tau}} \right) }Z_{\rm bos}(\tau,z)\,,
}
establishing (\ref{eq:PFModTrans}). Here we have normalized our currents to have level $k=1$. There is nothing special about our choice of free bosons, and indeed this transformation has also been worked out explicitly in other examples, for instance see \cite{Kiritsis:2007zza,AlvarezGaume:1986es} for free fermions, or \cite{Kawai:1993jk} for any chiral $\mathcal{N}=2$ theory.

\section{Bounds on the Charged-Spectrum Gap}
\label{sec:bounds}
In this section we derive bounds constraining what charged states must appear in a theory with a $U(1)$ global symmetry. We present constraints on what charges must appear, and upper bounds on the weight of the lightest charged state, and the ratio of weight to charge. 
\subsection{Hellerman-type Bound on Charged Spectrum Mass Gap}
It is immediately clear from the transformation property, (\ref{eq:CoordModTrans}), that there must be charged states in the theory. As a warm-up, it is worth writing down some simple bounds on what charges must show up. Setting $\bar{z}=0$ for simplicity, we can write the constraint of modular invariance as
\es{subtrans}{
0&=Z(-1/\tau, z/\tau ) - e^{\frac{i \pi z^2}{\tau}} Z(\tau, z)\\
 &=\sum_{i} \underbrace{\left(e^{2\pi i (Q_{i} z-h_{i})/\tau+2\pi i \bar{h}_{i}/\bar{\tau}}-e^{\frac{i \pi z^2}{\tau}}e^{2\pi i(Q_{i} z+h_{i}\tau-\bar{h}_{i}\bar{\tau})}\right)}_{F_{i}(\tau,\bar{\tau},z)}\,.
}
Here the sum is taken over individual states, each contributing $q^{h_{i}}\bar{q}^{\bar{h}_{i}}y^{Q_{i}}$, to the flavored partition function. Stronger constraints could be derived using the full Virasoro$\times U(1)$ characters, discussed in Appendix \ref{app:chartrans}. However we will get surprisingly strong results using the simpler single state expressions.

We can take $z$ derivatives of the modular relation, (\ref{subtrans}), to bring down factors of the charge $Q_{i}$ of each state, and then set $z$ to zero to obtain constraints on the charged spectrum. As a simple example, take two derivatives with respect to $z$ of the modular transformation equation (\ref{subtrans}), and evaluate at $z=0$, $\tau=i$. This gives
\es{statesum2}{
\frac{1}{8\pi^{2}}\sum_{i}\partial_{z}^{2}F_{i}\Big|_{z=0,\tau=i}&=\sum_{i}e^{-2\pi\Delta_{i}}\left(Q^{2}_{i}-\frac{1}{4\pi}\right)\,=\,0\,,
}
where $\Delta_{i}=h_{i}+\bar{h}_{i}$. This expression is negative for all $Q^{2}_{i}<1/4\pi$. In order for the sum to give zero, the theory must therefore have some states with $Q^{2}_{i}>1/4\pi$, in addition to the neutral states. One can do even better by taking more $z$ derivatives: combining the constraints from six and two derivatives, one finds,
\es{statesum6}{
\sum_{i}\left(\frac{1}{60 \pi ^3}\partial_{z}^6F_{i}\Big|_{z=0,\tau=i}-\frac{3}{2 \pi }\partial_{z}^{2}F_{i}\Big|_{z=0,\tau=i}\right)&=\sum_{i}e^{-2\pi\Delta_{i}}\left(\frac{32 \pi ^3 Q_{i}^6}{15}-8 \pi ^2 Q_{i}^4+1\right)=0\,.
}
This is positive for all $Q^{2}_{i}$, except for the interval, $0.344\lesssim Q_{i}\lesssim 1.09$. Thus the theory must have some states with charge in this range. 

\begin{figure}[t!]
\begin{center}
\includegraphics[width=0.47\textwidth]{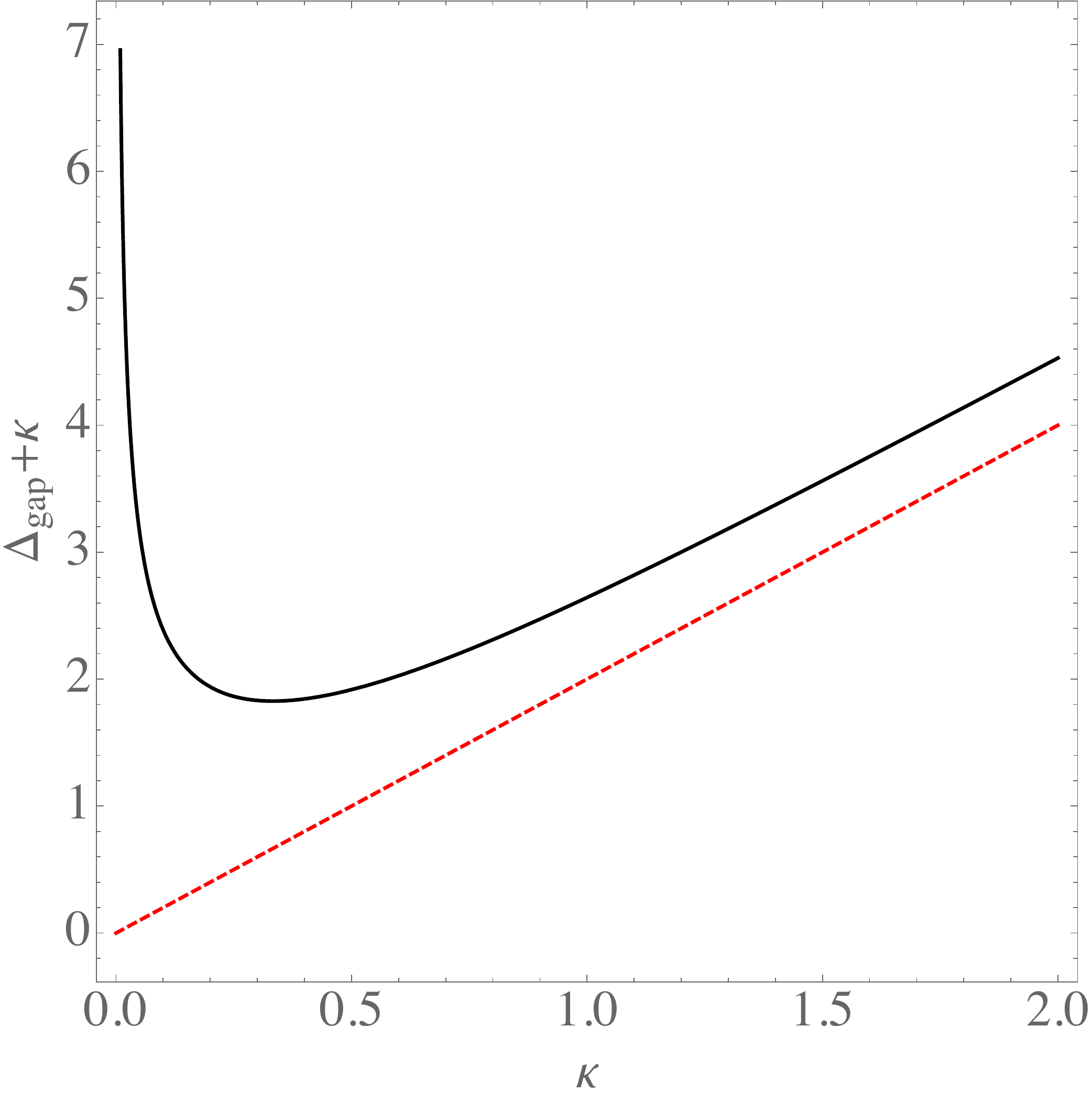}
\includegraphics[width=0.48\textwidth]{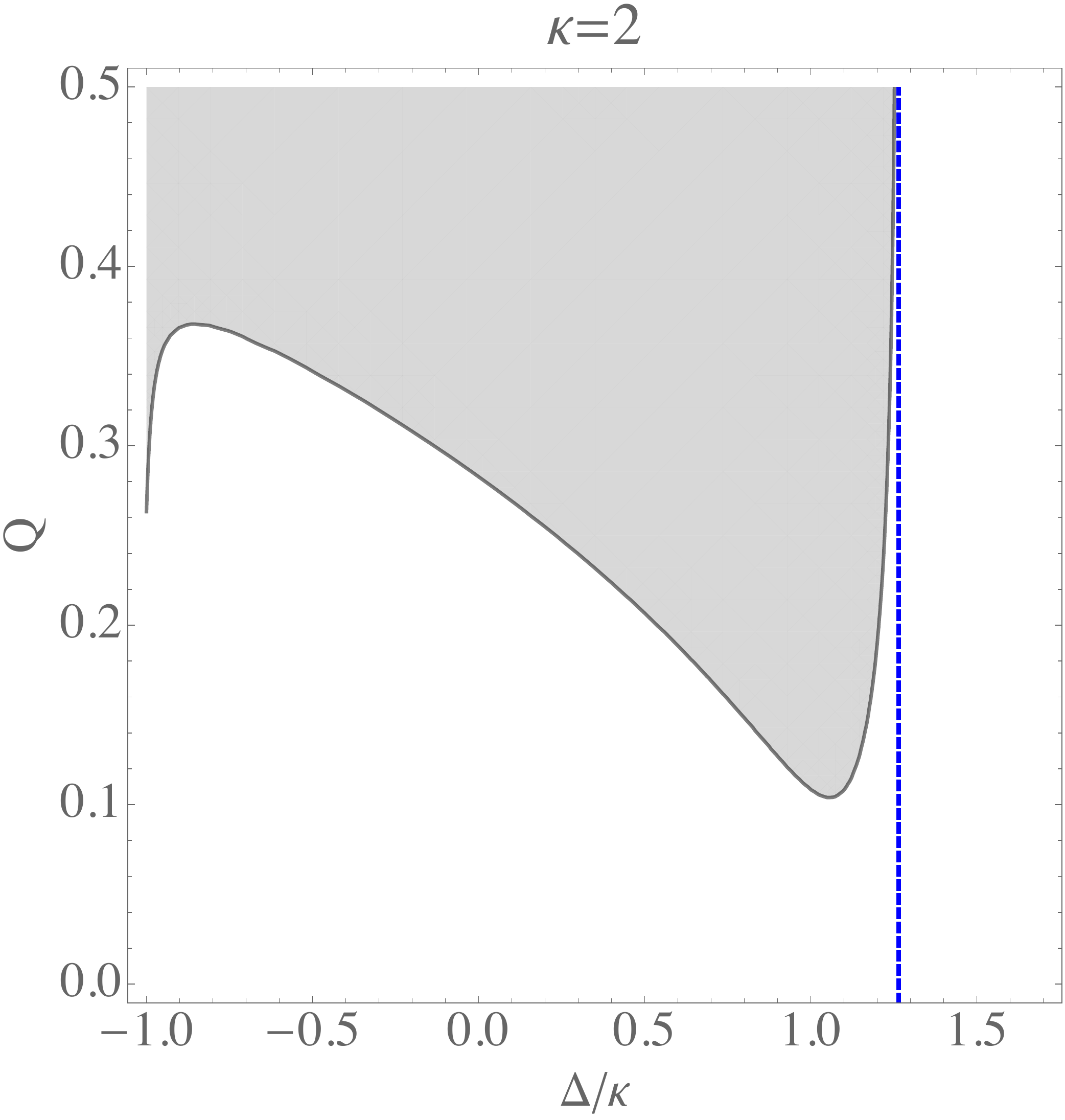}
\caption{{\it Left}: An upper bound on the total gap $\Delta_{\rm gap} + \kappa$ between the vacuum and the lightest charged state, as a function of $\kappa \equiv \frac{c+\bar{c}}{24}$. The slope asymptotes to $2\kappa$ at large $\kappa$, show in red, dashed. {\it Right}: 
The shaded region is where the polynomial $e^{2\pi \Delta} \alpha(F_{\Delta,Q})$ in (\ref{eq:examplealpha}) is negative for $\kappa=2$; the right edge asymptotes to a vertical line (shown in blue, dashed) at $\Delta/\kappa = \Delta_{\rm gap}(\kappa)/\kappa$ for large $Q$. In unitary theories, there must be at least one state in the shaded region.
}
\label{fig:gap}
\end{center}
\end{figure}

To prove a bound on the gap to the lightest charged state, our strategy (similarly to  most bootstrap approaches) will be to construct a linear operator $\alpha$ out of $z$ and $\tau$ derivatives evaluated at the self-dual point $z=0, \tau=i$, with the following properties:
\be
\alpha(F_{\rm vacuum}) = 1, &&  \nn\\
\alpha(F_{\Delta,Q}) > 0, && \textrm{ if } Q=0,\nn\\
\alpha(F_{\Delta,Q}) >0, && \textrm{ if } \Delta> \Delta_{\rm gap}.
\ee
Acting on the modular invariance equation (\ref{subtrans}), such an operator gives a positive contribution from the vacuum which must be canceled by a negative contribution from some states.  Since the only states that have $\alpha(F_{\Delta,Q})<0$ are charged states with $\Delta< \Delta_{\rm gap}$, it immediately implies that such states must be present in the theory.  This means $\Delta_{\rm gap}$ is an upper bound on the weight of the lightest charged state.\footnote{Incidentally, the unitarity bound $h+ \frac{c}{24} > \frac{Q^2}{2} $ means that any upper bound on the weight of a state is also an upper bound on its charge.}  An optimal analysis would seek to minimize $\Delta_{\rm gap}$ over the space of linear functionals subject to the above constraints on $\alpha$.  However, even with a small number of derivatives it is possible to obtain a functional $\alpha$ satisfying them.  Already quite non-trivial bounds are provided by the following example:
\be
\alpha(F_i) &\equiv& \left[ a_{1,0} \partial_\beta + a_{1,2} \partial_\beta \partial_z^2 + a_{3,0} \partial_\beta^3 + a_{3,2} \partial_\beta^3 \partial_z^2\right]F_i \left(\beta, z\right) \Big|_{z=0,\beta=2\pi}, \nn\\
a_{1,0} &=&\frac{1}{128} \left(-32 \pi ^3 \kappa ^3-64 \pi ^2 \kappa ^2-22 \pi  \kappa -128 \pi -3\right) , \nn\\
a_{1,2} &=&\frac{1}{64} \left(16 \pi ^2 \kappa ^3+24 \pi  \kappa ^2+13 \kappa +64\right) , \nn\\
a_{3,0} &=& \frac{1}{24} \pi^2 (3+\pi \kappa), \nn\\
a_{3,2} &=& - \frac{1}{24} \pi^2 \kappa, \ee
where we have taken $\tau = \frac{i \beta}{2\pi}, \bar{\tau} = -\frac{i \beta}{2\pi}$, and $\kappa \equiv \frac{c+\bar{c}}{24}$. Evaluated on the contribution $F_i$ from a single state, $\alpha$ produces the following polynomial:
\be
\alpha(F_{\Delta,Q}) &=& e^{-2 \pi \Delta} \left[ p_{0}(\Delta)+ Q^2 p_{1}(\Delta) \right] 
\label{eq:examplealpha}\nn\\
p_{0}(\Delta)&=&1+ (\Delta+\kappa) \frac{(3+4 \pi (\kappa-\Delta))^2}{64}\nn\\
 p_{1}(\Delta) &=& \pi ^3 \Delta ^2 \kappa -\frac{3}{2} \pi ^2 \Delta  \kappa -\frac{1}{16} \pi  \left(16 \pi ^2 \kappa ^3+24 \pi  \kappa ^2+5 \kappa +64\right).
 \ee
 At $Q=0$, this gives $p_{0}(\Delta)$ which is a manifestly positive polynomial for $\Delta \ge -\kappa$.  Furthermore, at sufficiently large $\Delta$, 
 \be
 p_{1}(\Delta) \approx \kappa \pi^3 \Delta^2 + {\cal O}(\Delta)
 \ee
 is also manifestly positive, so $\alpha(F_{\Delta, Q})$ is manifestly positive for all charged states as well when $\Delta$ is very large.  The only possible negative contributions come from charged states in the range of $\Delta$ where $p_{1}(\Delta) < 0$.  Thus, an upper bound on the gap is given by the larger of the two solutions to $p_{1}(\Delta)=0$:
 \be
 \Delta_{\rm gap}(\kappa) &=& \frac{\sqrt{2} \sqrt{\kappa  \left(8 \pi ^2 \kappa ^3+12 \pi  \kappa ^2+7 \kappa +32\right)}+3 \kappa }{4 \pi  \kappa } \nn\\
  &\approx & \kappa + \frac{3}{2\pi} + {\cal O}\left( \frac{1}{\kappa}\right).
 \ee
This is plotted as a function of $\kappa$ in Figure \ref{fig:gap}.  Also shown in Figure \ref{fig:gap} are contours of the polynomial $e^{2\pi \Delta} \alpha(F_{\Delta, Q})$ at $\kappa=2$, where one can see that the polynomial is negative only for non-zero $Q$ and for sufficiently small $\Delta$.  For a left-right symmetric theory, $\kappa = \frac{c}{12}$ and the vacuum is at $\Delta= - \frac{c}{12}$, so the bound on the gap between the lightest charge state and the vacuum is 
\be
\Delta_{\rm gap}(\kappa)-\Delta_{\rm vacuum} \approx \frac{c}{6}  + \frac{3}{2\pi} + {\cal O}\left( \frac{1}{c} \right).
\ee
More restrictive bounds can certainly be obtained by considering more derivatives of $F_i$ than we have used here, and it would be interesting to explore the optimal bounds that can be obtained this way.

\subsection{Bounds on Charge-to-Mass Ratio}
So far we have investigated the bounds on the gap in charge, and the gap in the weight of the lightest charged state. It is interesting to also ask what we can say about a maximal gap in the ratio of weight to charge. For fixed central charge, the operator $\alpha$ defined in the previous section already provides a bound on this ratio, as for large enough $\Delta$ or small enough $Q$, $\alpha(F_{\Delta,Q})>0$. We will be most interested, however, in obtaining a bound for large $c$.\footnote{We often think about theories with a given level and a quantized, order one $U(1)$ charge. In these cases, our bound on the weight of the lightest charged state immediately translates into a bound on the ratio, and gives a bound that scales with the central charge. By applying the linear operator techniques of the previous subsection, we will be able to derive a similar bound,  that holds more generally without any additional assumptions on quantization.}

To this end, define a new linear functional $\tilde{\alpha}$ as
\es{newop}{
\tilde{\alpha}(F_{i})&=\alpha(F_{i})+\frac{\pi}{4}\kappa^{3}\partial_{z}^{2}F_{i}(\beta,z)\Big|_{z=0,\beta=2\pi}\,.
}
Acting on a single state, this again produces $e^{-2 \pi \Delta}$ times a relatively simple polynomial:
\es{newopact}{
\tilde{\alpha}(F_{\Delta,Q})&=e^{-2\pi\Delta}\left[\tilde{p}_{0}(\Delta)+Q^{2}\tilde{p}_{1}(\Delta)\right] ,\\
\tilde{p}_{0}(\Delta)&=p_{0}(\Delta)+\frac{\pi^{2}}{2}\kappa^{3}, \\
\tilde{p}_{1}(\Delta)&=p_{1}(\Delta)-2\pi^{3}\kappa^{3}\,.
}
As before, we want to investigate for what states these polynomials can be negative, focusing on how the mass-to-charge bounds scale in the large central charge limit.   We can therefore look mainly at $\kappa$ large, and divide up our analysis into the three regimes $\Delta \gg \kappa, \Delta \ll \kappa$, and $\Delta/\kappa \sim {\cal O}(1)$.

For large $\Delta$, $\Delta\gg\kappa$, we have,
\es{largedim}{
e^{2\pi\Delta}\tilde{\alpha}(F_{\Delta,Q})\approx\frac{\pi^{2}}{4}\Delta^{3}+\pi^{3}\kappa Q^{2}\Delta^{2}\,,
}
which is positive for all states.

For small $\Delta$, that is $\Delta-\Delta_{\rm vacuum}\ll\kappa$, we have,
\es{smalldim}{
e^{2\pi\Delta}\tilde{\alpha}(F_{\Delta,Q})\approx 2\pi^{3}\kappa^{3}\left(\frac{1}{4\pi}-Q^{2}\right)\,.
}
This is negative only for states with $Q^{2}>1/4\pi$, and thus such states have a very small mass-to-charge ratio $(\Delta-\Delta_{\rm vacuum})/Q\ll\kappa$. 

The most interesting states are those with $\Delta-\Delta_{\rm vacuum}\sim \kappa$. In this case,
\es{cdim}{
e^{2\pi\Delta}\tilde{\alpha}(F_{\Delta,Q})\approx \frac{\pi ^2}{4} \left((\kappa -\Delta )^2 (\Delta +\kappa )+2\kappa ^3\right)+Q^{2}\pi ^3 \kappa  \left(\Delta ^2-3 \kappa ^2\right)\,.
}
The $Q$ independent term is again positive, while the second term can be negative for sufficiently small $\Delta$. For the total expression to be negative we must have,
\es{Qbound}{
Q^{2}\geq\frac{ \left((\kappa -\Delta )^2 (\Delta +\kappa )+2\kappa ^3\right)}{4 \pi\left(\kappa  \left(3 \kappa ^2-\Delta ^2\right)\right)}\,, \ \ \ \ \ \Delta\in\left(-\kappa,\kappa\sqrt{3}\right).
} 
Though not uniform in $\Delta$, the right hand side of (\ref{Qbound}) has a minimum in the allowed range of $\Delta$. The quantity of interest is thus bounded by,
\es{deltaoverQ}{
\frac{\Delta-\Delta_{\rm vacuum}}{|Q|}\leq\frac{\Delta-\Delta_{\rm vacuum}}{\sqrt{\frac{ \left((\kappa -\Delta )^2 (\Delta +\kappa )+2\kappa ^3\right)}{4 \pi\left(\kappa  \left(3 \kappa ^2-\Delta ^2\right)\right)}}}\leq 4\sqrt{\pi}\kappa\,.
}
We see that the largest gap in weight per unit charge scales linearly with the central charge.

A few comments are in order concerning the linear in $c$ scaling of the bound and the connection with the WGC.  In most previous work on the WGC, a quick way to read off how a  bound ought to depend parametrically on $G_N$ is to use dimensional analysis to fix the power $a$ of $G_N$ in the dimensionless quantity $\frac{m}{g Q G_N^a}$, where $g$ is the gauge coupling.  In $d=3$, $gQ$ has mass dimension $1/2$, for a Maxwell gauge field, and as a result the bound scales like $G_N^{-1/2}$ rather than $G_N^{-1}$.  However, one should remember that a boundary current in 2d is dual to a Chern-Simons gauge field in AdS$_3$, rather than a gauge field with just a Maxwell kinetic term $-\frac{1}{4 g^2} F^2$.  Since the CS coupling is dimensionless, in our case naive dimensional analysis would actually suggest a bound that scales linearly in $G_N^{-1}$, which agrees with what we find.  It would be interesting to make this connection sharper, by adapting previous arguments to the case where the kinetic term of the 3d gauge field is dominantly CS.  

The fact that CS terms are more relevant than Maxwell kinetic terms suggests that the former may be pertinent for a wider range of theories than the latter are; in any case they are more directly connected to the dynamics of boundary conserved currents in 2d and therefore more amenable to bootstrap bounds.  It would therefore be interesting to extend the analysis of \cite{Heidenreich:2015nta} to include an investigation of what role CS terms play after compactification from higher dimensions down to 3d.  

Finally, one could also hope to study a regime where the Maxwell term is present in the bulk and sufficiently large that it dominates the CS term 
over a hierarchy of energy scales below the Planck scale. Previous arguments for a WGC suggest that a stronger bound on the mass-to-charge ratio, scaling like $G_N^{-1/2}$, should emerge.   This regime of theories would correspond to those with a ``weakly coupled'' bulk gauge field, whose mass would be parametrically small compared to the Planck scale, and perhaps by assuming the presence of the dual operator in the CFT one could obtain much stronger bounds on the spectrum of charged states with the modular bootstrap.

\subsection{Bound from Asymptotic Growth}
We next present an alternate method for deriving a bound on the gap in the charged sector, subject to a non-cancellation hypothesis. Although this argument will require a mild extra assumption, the advantage is both that it is very simple, and it is similar in style to an argument we will use to derive stronger bounds for $\mathcal{N}=(1,1)$ theories. This method is more analogous to the original argument due to Cardy for the asymptotic density of states in a 2d CFT.  More accurately, it is analogous to the inverse of Cardy's argument; rather than using the presence of the vacuum to ascertain the asymptotic growth of states at large $\Delta$, we will show that a non-vanishing asymptotic charge density implies the presence of a light charged state.  We will argue for this by considering the following object,
\es{w4def}{
W_{4}(\tau,\bar{\tau})&=Z(\tau,\bar{\tau})\partial_{z}^{4}Z(\tau,\bar{\tau})\Big|_{z=0}-3 (\partial_{z}^{2}Z(\tau))^{2}\Big|_{z=0}\,.
}
Note that this function vanishes if there are no charged states. Using the modular transformation properties of the flavored partition function  (\ref{eq:flavorflav}), it follows that $W_{4}$ transforms as
\es{w4trans}{
W_{4}\left(\tau^{\prime},\bar{\tau}^{\prime}\right)&=(c\tau+d)^{4}W_{4}(\tau,\bar{\tau})\,.
}
Considering $W_{4}$ for imaginary $\tau=i\beta/2\pi$, we have,
\es{W4exp}{
W_{4}(\beta)&=\frac{1}{2}\sum_{i,j}\left(Q_{i}^{4}-6Q_{i}^{2}Q_{j}^{2}+Q_{j}^{4}\right)e^{-\beta(\Delta_{i}+\Delta_{j})}\\
&=\sum_{\tilde{\Delta}}C_{\tilde{\Delta}}e^{-\beta \tilde{\Delta}}\,.
}
In the last line we have written the sum over weights, $\tilde{\Delta}=\Delta_{i}+\Delta_{j}$. 

Assume for contradiction that the first charged state has weight $\Delta_{\rm gap}$. Then the sum in $W_{4}$ starts at $\tilde{\Delta}=\Delta_{\rm gap}-c/12$. We will show that this is inconsistent for large enough $\Delta_{\rm gap}$.

In order to see this, it is instructive to consider an abstract function which is invariant under the real modular $S$ transformation,
\es{weight0}{
W_{0}\left(\frac{4\pi^{2}}{\beta}\right)\,=\,W_{0}(\beta)\,=\,\sum_{\tilde{\Delta}\geq\Delta_{0,{\rm gap}}-c/12}D_{\tilde{\Delta}}e^{-\beta \tilde{\Delta}}\,.
}
To make contact with $W_{4}$ above, we will assume $W_{0}$ has $|D_{\tilde{\Delta}}|$ growing with large $\tilde{\Delta}$.\footnote{This assumption is tantamount to an asymptotic non-cancellation assumption between the combination of partition functions appearing in (\ref{w4def}). In fact, the situation is even better: even if this particular combination had a cancellation, we could construct higher order modular objects, and rerun the argument using these higher order modular forms.  In this case, a different particular combination would have to cancel to invalidate the argument.  Obviously, we could repeat this as many times as needed until reaching a combination that did not cancel. Thus to invalidate this argument would require an infinite number of cancellations.} If we further take $\Delta_{0,{\rm gap}}>c/12$, then,
\es{limits}{
\lim_{\beta\rightarrow\infty}W_{0}(\beta)\,=\,\lim_{\beta\rightarrow0}W_{0}(\beta)\,=\,0\,.
}
Thinking of $W_{0}(\beta)$ as the Laplace transform of $D_{\tilde{\Delta}}$, the final value theorem\cite{RASOF1962165} tells us that the large $\tilde{\Delta}$ behavior of $D_{\tilde{\Delta}}$ is given by the small $\beta$ behavior of $W_{0}(\beta)$, and thus $\lim_{\tilde{\Delta}\rightarrow\infty}D_{\tilde{\Delta}}=0$, contradicting our assumed growth. This tells us we must have $\Delta_{0,{\rm gap}}-\Delta_{\rm vacuum}\leq c/6$.

To apply this to the function, $W_{4}$, that we are interested in, we divide by the modular discriminant, ${\bf\Delta}(\beta)=\eta^{24}(i\beta/2\pi)$ to the appropriate power to create an invariant function:\footnote{In running this argument, it is crucial that ${\bf\Delta}$, not to be confused with the weight $\Delta$, only has zeroes at the cusp $\beta\rightarrow 0 \sim \infty$, as otherwise we would introduce extra poles in $\hat{W}_{4}$, invalidating the applicability of the final value theorem.}
\es{normw4}{
\hat{W}_{4}(\beta)&=\frac{W_{4}(\beta)}{({\bf\Delta}(\beta))^{1/3}}\\
\hat{W}_{4}\left(\frac{4\pi^{2}}{\beta}\right)&=\hat{W}_{4}(\beta)\,.
}
The above argument tells us that $\hat{W}_{4}$ has to grow as $\beta\rightarrow\infty$. Since the modular discriminant behaves as $(\mathbf{\Delta}(\beta))^{1/3}\sim e^{-\beta/3}$, we must have a maximal gap to charged states of
\es{chargegap}{
\Delta_{\rm gap}-\Delta_{\rm vacuum}=c/6+1/3\,.
}

\subsection{Supersymmetry}

As we have mentioned, we don't believe the bound ($\ref{chargegap}$) is optimal. One motivation for this conjecture comes from considering theories with additional symmetry. We can consider the case of a 2d CFT with both $\mc{N}=(1,1)$ supersymmetry and a $U(1)$ current. With $\mc{N}=(1,1)$ supersymmetry we can define a holomorphic quantity called the elliptic genus. This theory has fermions, so when we put it on a torus, there are four different spin structures we can consider depending on boundary conditions. We thus define the following elliptic genera,
\begin{align}
Z_{\text R}^+(\tau,z) &= \Tr_{\text{R,R}}((-1)^{F_R}q^{L_0-\frac{c}{24}} y^{J_0} \bar{q}^{\bar{L_0}-\frac{\bar{c}}{24}}), \nn \\
Z_{\text R}^-(\tau,z) &= \Tr_{\text{R,R}}((-1)^{F_L + F_R}q^{L_0-\frac{c}{24}} y^{J_0} \bar{q}^{\bar{L_0}-\frac{\bar{c}}{24}}), \nn \\
Z_{\text{NS}}^+(\tau,z) &= \Tr_{\text{NS,R}}((-1)^{F_R}q^{L_0-\frac{c}{24}} y^{J_0} \bar{q}^{\bar{L_0}-\frac{\bar{c}}{24}}), \nn \\
Z_{\text{NS}}^-(\tau,z) &= \Tr_{\text{NS,R}}((-1)^{F_L+F_R}q^{L_0-\frac{c}{24}} y^{J_0} \bar{q}^{\bar{L_0}-\frac{\bar{c}}{24}}).
\label{eq:warpig}
\end{align}
In all of the functions above, the right-moving sector  gets contributions  only from supersymmetric ground states at $\bar{L}_0=\frac{\bar{c}}{24}$.\footnote{Note, here the left-moving fermion number for the NS vacuum is conventionally defined as $(-1)^{c/6}$.} The advantage of considering the elliptic genus is that it is a holomorphic modular form, so we can use the power of holomorphy to bound the gap to the lightest charge state. 

The functions in (\ref{eq:warpig}) transform as (\ref{eq:PFModTrans}) under (some congruence subgroup of) $SL(2,\bb{Z})$. In particular, the functions $Z_{\text R}^+(\tau,z), Z_{\text R}^-(\tau,z), Z_{\text{NS}}^+(\tau,z),$ and $Z_{\text{NS}}^-(\tau,z)$ transforms as (\ref{eq:PFModTrans}) under $\Gamma_0(2), SL(2,\bb{Z}), \Gamma^0(2)$, and $\Gamma_\th$ respectively. These are defined as
\begin{align}
\Gamma_0(2) &\equiv \left\{\left(\begin{array}{cc}a&b\\c&d\end{array}\right) \in SL(2,\bb{Z}),~~c \equiv 0 ~({\rm mod} ~2)~\right\} , \nn \\
\Gamma^0(2) &\equiv \left\{\left(\begin{array}{cc}a&b\\c&d\end{array}\right) \in SL(2,\bb{Z}),~~b \equiv 0 ~({\rm mod} ~2)~\right\} , \nn \\
\Gamma_\th &\equiv \left\{\left(\begin{array}{cc}a&b\\c&d\end{array}\right) \in SL(2,\bb{Z}),~~a+b \equiv 1 ~({\rm mod} ~2), ~c+d \equiv 1~({\rm mod}~2)~\right\}.
\end{align}
These functions transform into each other via
\begin{align}
Z_{\text{R}}^+(\tau, z) &= Z_{\text{NS}}^-(-1/\tau, z /\tau), \nn \\
Z_{\text{NS}}^-(\tau, z) &= e^{-\frac{2\pi i c}{24}}Z_{\text{NS}}^+(\tau+1, z).
\label{eq:wardogs}
\end{align}

Now let us consider the following function:
\es{mod4susy}{
W_{4}^{\text{R}}(\tau)\equiv Z_{\text R}^+(\tau,z)\partial_{z}^{4}Z_{\text R}^+(\tau,z)\Big|_{z=0}-3 (\partial_{z}^{2}Z_{\text R}^+(\tau,z))^{2}\Big|_{z=0}\,.
}
This is a weight 4 modular form under $\Gamma_0(2)$. Moreover, the only contributions to $W_4^{\text{R}}(\tau)$ come from charged states. Our basic strategy is to show that $W_4^{\text{R}}(\tau)$ must have a term of at least $\mc{O}(q)$ when expanded about $\tau=i\infty$; this then means that there must be at least one charged state of dimension one above the RR vacuum. Thus, relative to the NS-NS vacuum, we must have a charged state by $\frac{c}{12}+1$.

The ring of modular forms under $\Gamma_0(2)$ is generated by the functions $E_2'(\tau)$ and $E_4(\tau)$, defined in Appendix \ref{app:modulardefinitions}. In particular, any meromorphic function that transforms with weight $w$ under $\Gamma_0(2)$ that has no poles at $\tau=i\infty$ and diverges at most as $\tau^{-w}$ about $\tau=0$ can be written as a linear combination of products of $E_2'$ and $E_4$ \cite{ModularStein}.

To see that $W_4^{\text{R}}$ is a weight four modular form under $\Gamma_0(2)$, note that about $\tau=i\infty$, $W_4^{\text{R}}$ is finite, as the lightest Ramond sector states have weight zero. The only question is the behavior about $\tau=0$. 

Suppose we have a theory with the first charged state at least $\frac{c}{12}$ above the (NS-NS) vacuum. From (\ref{eq:wardogs}) and (\ref{mod4susy}), one can show
\be
W_4^{\text R}(\tau) = \frac1{\tau^4} W_4^{\text{NS}}\(-\frac1\tau\),
\label{eq:toad}
\ee
where we define
\be
W_4^{\text{NS}}(\tau) \equiv Z_{\text{NS}}^-(\tau,z)\partial_{z}^{4}Z_{\text{NS}}^-(\tau,z)\Big|_{z=0}-3 (\partial_{z}^{2}Z_{\text{NS}}^-(\tau,z))^{2}\Big|_{z=0}\,.
\label{eq:toadmahajan}
\ee
Note that $W_4^{\text{NS}}(\tau)$ also only gets contributions from charged states. In particular, as we've assumed the first charged state shows up $\frac{c}{12}$ above the vacuum, then $W_4^{\text{NS}}(\tau)$ has no poles about $\tau=i\infty$. Thus using (\ref{eq:toad}), we see that $W_4^{\text{R}}(\tau)$ diverges at most as $\tau^{-4}$ as $\tau=0$. This means it can be written as
\be
W_4^{\text R}(\tau) = c_1 E_2'(\tau)^2 + c_2 E_4(\tau)\,,
\label{eq:toadraghu}
\ee
for some constants, $c_{1}$ and $c_{2}$.

The highest order in the $q$-expansion (\ref{eq:toadraghu}) can start at is $\mc{O}(q)$. Thus, in $W_4^{\text{R}}$, a charged state must appear by dimension at least one above the RR vacuum. Since the RR vacuum is $\frac{c}{12}$ above the NS-NS vacuum, we thus get a bound to the first charged state of
\be
\Delta-\Delta_{\text{vacuum}}\leq\frac{c}{12}+1 \qquad \textrm{(supersymmetric})\,.
\ee

The improvement by a factor of 2 compared to our non-supersymmetric bounds brings this into line with the threshold for BTZ black holes, since dimensions of $\Delta-\Delta_{\rm vac} \sim \frac{c}{12}$ correspond to masses $m \sim \frac{1}{8 G_{\rm N}}$ in the gravity picture. 
It seems natural to conjecture that  a bound upper bound on charged states of order $\sim\frac{c}{12}$ may hold in general, even in the non-supersymmetric case.  

\section{Large Gap Examples}
\label{sec:largegapex}

In this section we provide some examples of theories which realize our bound up to $O(1)$ factors. 
 One class of examples is given by free bosons compactified on extremal lattices. Such lattices can be explicitly constructed for small central charge and are known not to exist for $c\geq163264$ \cite{JenkinsRouse}. Appealing to more standard string theory examples, we also consider a gravitational theory in flat space, and discuss  the D1-D5 system in highly curved AdS space. 
\subsection{Extremal Lattices}\label{sec:extlat}

 An extremal lattice, $\Lambda_{c}$, is a rank $c$ even self dual lattice with the smallest norm non-zero vector, $\vec{v}^{*}$ having length squared,
 \es{smallnorm}{
\vec{v}^{*} \cdot\vec{v}^{*}&=\frac{c}{12}+2\,.
 }
We will be focused on the case $c=24k$ for $k\in\mathbb{Z}$. Such lattices are known to exist for $k=1$,$2$, and $3$ \cite{Leech:1967,Nebe1998472,Nebe:2010}, however for larger $k$ they have not been constructed. As mentioned above, they do not exist for sufficiently large $k$, $k>6802$.

A consistent chiral CFT can be constructed by considering $c$ chiral bosons compactified on such a lattice \cite{Caselle:1988jm,Dolan:1989vr}. This CFT has a spectrum consisting of the vertex operators,
\es{vert}{
\mathcal{V}_{\vec{v}}(z)&=e^{i\vec{v}\cdot\vec{\phi}}(z)\,, \ \vec{v}\in\Lambda_{c} \ \ \ \ h_{\vec{v}}\,=\, \frac{\vec{v}^{2}}{2}\,,
}
as well as the differentials, $-i\partial\vec{\phi},\,\partial\vec{\phi}^{2},\ldots$.

The differentials, $-i\partial\vec{\phi}$, form a set of $c$ currents, under which the only charged operators are the vertex operators, $\mathcal{V}_{\vec{v}}$\,. Consider any one of these currents,
\es{extlatcur}{
J(z)&=-i\partial\phi^{1}(z)\,.
}
The gap to the first charged operator is given by the gap in the norm of vectors in $\Lambda_{c}$, and thus,\footnote{It can actually be shown that chiral CFTs satisfy a stricter bound on the weight of the lightest charged state, $h_{\rm gap}-h_{\rm vacuum}\leq\frac{c}{24}+1$, and so these examples are tight for chiral CFTs \cite{amwip}.}
\es{}{
h_{\Lambda_{c},\,{\rm gap}}-h_{\rm vacuum}&=\frac{c}{24}+1\,.
}
\subsection{Gravity theories with a large gap}

It is expected for a variety of reasons that quantum gravity theories with $U(1)$ gauge fields will exhibit charged
matter with charge of ${\cal O}(1)$ at a mass scale $M \lesssim M_{\rm Planck}$.  As 2d CFTs are (sometimes) dual 
to weakly curved 3d gravity, one can ask: how does our bound compare to this expectation?

In light of the Brown-Henneaux formula 
\es{bh}{c = {3 {L}_{AdS} \over 2{ G}}\,,}
our bound is sufficient to guarantee this expectation.  Charged states at masses $M \sim {c\over 6}$ in
AdS units (the highest value consistent with the bound), are at a mass $\sim M_{\rm Planck}$.
Still, one might wonder -- is a stronger absolute bound possible in weakly curved gravitational theories?

We think the answer is no.  One can easily provide examples of gravity theories which are thought to be fully
consistent, yet have abelian gauge fields with the first charges appearing at $\sim M_{\rm Planck}$.  We provide
two examples below.  It is important to stress that in each, our ability to make controlled statements depends 
on extended supersymmetry and exact BPS mass formulae, as we work in regimes where some size or coupling is
of ${\cal O}(1)$.

\bigskip
\noindent
{\bf Example 1}: 

\noindent
Consider M-theory compactified on a circle of radius $R$ in 11d Planck units.  At very large radius,
the theory reduces to 11d supergravity.  At very small radius, one can reinterpret the radius in terms of the type IIA
string coupling, via
\es{rform}{R = g_{\rm string}^{2/3}\,.}For any finite $R$, the long distance theory is a weakly curved gravity theory in ten dimensions.  

There is a Kaluza-Klein gauge field arising from the $\mu 11$ components of the 11d metric.  This gauge field becomes the Ramond-Ramond photon of type IIA string theory as $R \to 0$.  But it is present for all values of $R$, and a BPS
bound relates the mass of the lightest charged KK modes of a given charge to the radius of the circle.  

Half-BPS states carrying this charge do exist.  They are the Kaluza-Klein gravitons on the circle, or D0-brane bound states in the IIA string.  When $R = {\ell}_{11}$, the only mass scale in the BPS formula is $M_{\rm Planck,11}$, and
the lightest charge has mass $\sim M_{\rm Planck,11}$.

At long distances, one then has gravity coupled to an abelian gauge field in 10d flat space, with a lightest charge
at $M_{\rm Planck,10} \sim M_{\rm Planck,11}$. This easily generalizes to lower dimensions, by compactifying on a Planck radius torus, rather than a single circle.
This shows that one cannot derive a stronger bound on the mass of the lightest charged state which is stronger than
the Planckian bound, at least not one which applies to all weakly curved gravity theories.\footnote{In fact, in 10d we can shrink the circle, thereby taking $g_{\rm string}$ small, and the only charged states in the theory are D0 branes, which remain above the Planck scale.  This provides an example at small string coupling; however, the gauge coupling remains ${\cal O}(1)$. }

\bigskip
\noindent
{\bf Example 2}:

\noindent
Our bound is more directly related to AdS$_3$ gravity theories, via the relationship between large $c$ 2d CFTs with sparse spectrum and weakly curved gravity.  So one could ask -- in that more limited context, could it be that there is
a (parametrically) stronger bound available?

We will try to give some sense of whether a counter-example may or may not exist by discussing one canonical example of AdS$_3$/CFT$_2$.
Unfortunately, this example comes close to our bound only at small AdS length and thus at small $c$, whereas what we want to compare to is the parametric dependence on the bound at large $c$. The problem of finding weakly curved AdS$_{3}$ examples with a large gap to charged states is similar to the problem of constructing very sparse large $c$ CFTs and is likely challenging. However, at present it is unclear whether this is a fundamental limit, or just a limitation of available controlled compactifications methods.

So, let us discuss the original example of AdS$_3$/CFT$_2$ duality, coming
from the D1-D5 system on $T^4$.  Before inserting the branes and taking the near-horizon limit, the moduli space of compactifications of type IIB string theory on $T^4$ is a coset space
\es{modspace}{SO(5,5;{\mathbb Z})~\backslash ~SO(5,5)~ \slash ~SO(5) \times SO(5)\,.}
Inserting the $Q_1$ $D1$ and $Q_5$ (wrapped) $D5$ branes leaves a worldvolume unbroken (4,4) supersymmetric theory on the black string in six dimensions.  The
25 real moduli can be divided into background tensor multiplet and hypermultiplet scalars of this supersymmetry; 5 come from tensor multiplets and 20 from hypermultiplets.

Via the attractor mechanism, the tensor multiplet scalars take fixed values in the near-horizon geometry, independent of our choices.  The hypermultiplet scalars can be tuned at will.

The resulting near-horizon solution is
\es{nearh}{\left({\rm AdS}_3 \times {\rm S}^3\right)_{Q_1 Q_5} \times {\rm T}^4~.}
The radius of the AdS space and the sphere are equal (as is standard in Freund-Rubin compactification), given by
\es{fr}{R_{\rm AdS}^2 = \alpha^\prime g_6 \sqrt{Q_1 Q_5}~.}

The two moduli of significance for us are the 6d string
coupling $g_6$, and the $T^4$ volume $v$.  In string
units, the volume is given by
\es{vol}{v = {Q_{1} \over Q_5}~,}
while $g_6$ is in a hypermultiplet and we are free to choose its value.
Validity of the 6d supergravity description requires weak AdS curvature, i.e. 
\es{sg}{g_6 \sqrt{Q_1 Q_5} \gg 1~.}

Consider, then, the scaling limit
\es{scale}{Q_1 \to \infty, ~Q_5 \to \infty,  ~{Q_1 \over Q_5} \sim 1}
while simultaneously selecting
\es{scale2}{g \lesssim {\cal O}(1)~.}
In this limit, the 6d supergravity theory is weakly
curved, while $v \sim {\cal O}(1)$.  So the 6d string
and Planck scales are comparable.

Now, consider the KK U(1) gauge fields on the torus.  
The story is similar to that of Example 1; the lightest
charges will be KK modes with six-dimensional masses $\sim M_{\rm Planck,6}$. To read off the AdS$_{3}$ mass, we need to further reduce on the S$^{3}$. Unfortunately, as the AdS and sphere radius are tied, the three dimensional mass is well below the Planck mass.

We can produce theories where the lightest charged states are at the Planck mass in this example, but only by considering highly curved theories outside of the supergravity limit, (\ref{sg}), by taking $R_{\rm AdS}=R_{\rm S^{3}}=\mathcal{O}(1)$ in Planck units.
It is clear that the problem is that the Freund-Rubin construction by definition ties the AdS radius to the radius of an external sphere in the geometry.  So at large AdS radius, the dilution of the lower-dimensional (AdS) Planck scale due to the external sphere, will always lower the gap to charges under a KK gauge field.  More elaborate constructions can partially surmount this issue, but we are not aware of any where we would calculably saturate our bound at large AdS radius.

\section{Discussion and Future Directions}
\label{sec:disc}

We have demonstrated that the partition function with a chemical potential can be used to put concrete bounds on the spectrum of charged states in a general, not necessarily holographic, 2d CFT. Interpreted in terms of gravitational duals, these imply that charged states must be present in the theory at the Planck scale or lower, and that furthermore there must exist states with charge-to-mass ratio (in units of the Planck scale) above a concrete lower bound. For the most part, we have attempted to make our analysis more analytically transparent at the cost of leaving the constraints weaker than should ultimately be possible, and it would be interesting to return to these bounds with the much more numerically sophisticated machinery of recent bootstrap approaches.\footnote{See e.g. \cite{Rattazzi:2008pe,Friedan:2013cba,Poland:2011ey,ElShowk:2012ht,ElShowk:2012hu,Kos:2014bka}, to name just a few of the many such analyses in this rapidly growing area.  } 

We also expect that these methods could be generalized to bound other quantities besides those considered here.  For one, we have focused only a single conserved current, but when its symmetry is part of a larger non-abelian group, then one should be able to make richer statements about the spectrum of charges.  In particular, rather than simply bounding the charge $Q$ of states, one could start to constrain the representations of states in the theory.  It would be very interesting for instance to show that for certain symmetry groups, certain representations must appear in the spectrum, or to find relations between the representations that appear in the low-energy spectrum with those at high energies.

Another potentially powerful extension would be to correlation functions in higher dimensions.  This paper has focused on the partition function, but in two-dimensional CFTs this is equivalent to a four-point correlation function of twist operators \cite{Witten:2007kt}. Adding in a chemical potential is equivalent to inserting Wilson lines in this correlation function. Optimistically, one may hope that even in this more general case, the transformation property of the correlator under crossing in the presence of such Wilson lines can be derived purely through knowledge of the current two-point function, or in even dimensions in terms of its anomalies.\footnote{See for instance \cite{MooreLectures}, section 3.1.4 for a very rough sketch of such an argument in $d=2$.}  If this is correct, then it would provide a practical way of including non-local line operators in the conformal bootstrap, potentially accessing important information about the theory that would be invisible otherwise \cite{Aharony:2013hda}. 

Finally, bounds on the number of BPS operators at a given weight and charge in a 2d superconformal field theory with at least ${\cal N}=2$ supersymmetry are of additional interest, as they would have a topological interpretation as bounds on the Hodge numbers of the corresponding target-space K\"ahler manifold.\footnote{See \cite{Keller:2012mr, Fiset:2015pta} for various approaches to this question.} Such constraints are therefore interesting geometrically, and modular bootstrap approaches may provide information that is complementary to other approaches.



\section*{Acknowledgments}

We thank Andy Cohen, Thomas Dumitrescu, Guy Gur-Ari, Dan Harlow, Daniel Jafferis, Jared Kaplan, Ami Katz, Alex Maloney, Greg Moore, Eric Perlmutter, Riccardo Rattazzi, Matthew Reece, Cumrun Vafa, Herman Verlinde, Roberto Volpato, and Xi Yin for useful discussions.  We also thank Dan Harlow, Ami Katz, Christoph Keller, and Alex Maloney for comments on a draft. NB is supported by a Stanford Graduate Fellowship and an NSF Graduate Fellowship. ED is supported by the NSF under grant PHY-0756174. ALF is supported by the US Department of Energy Office of Science under Award Number DE-SC-0010025.  SK acknowledges the support of the National Science Foundation via grant PHY-1316699.

\begin{appendices}

\section{Modular Forms}
\label{app:modulardefinitions}
For convenience, we reproduce the definitions and relevant properties of several functions used in this paper. The Eisenstein series $E_4(\tau)$ and $E_6(\tau)$ are defined as
\begin{align}
E_4(\tau) &= 1 + 240\sum_{n=1}^{\infty} \frac{n^3q^n}{1-q^n} \nn \\
E_6(\tau) &= 1- 504\sum_{n=1}^{\infty} \frac{n^5q^n}{1-q^n}
\label{eq:e4blahblah}
\end{align}
They transform as
\begin{align}
E_4\(\frac{a\tau+b}{c\tau+d}\) &= (c\tau+d)^4E_4(\tau) \nn \\
E_6\(\frac{a\tau+b}{c\tau+d}\) &= (c\tau+d)^6E_6(\tau). 
\end{align}
Together, they generate the ring of modular forms invariant under $SL(2,\bb{Z})$. We also define the Dedekind eta function as
\be
\eta(\tau) = q^{\frac{1}{24}} \prod_{n=1}^{\infty} (1-q^n)
\ee
and the modular discriminant as
\be
{\bf \Delta}(\tau) = \eta(\tau)^{24} = \frac{E_4(\tau)^3-E_6(\tau)^2}{1728}.
\ee
We are also occasionally interested in the second Eisenstein series $E_2(\tau)$, defined as
\be
E_2(\tau) = 1-24\sum_{n=1}^{\infty} \frac{n q^n}{1-q^n}.
\ee
This is not quite a modular form, as it transforms as
\be
E_2\(\frac{a\tau+b}{c\tau+d}\) = (c\tau+d)^2E_2(\tau) + \frac{6c}{i\pi}(c\tau+d).
\label{eq:e2trans}
\ee
We also define the Klein-invariant $J$ function, which is a modular function of weight 0 with a pole at $\tau=i \infty$.
\be
J(\tau) = \frac{E_4(\tau)^3}{{\bf \Delta}(\tau)}-744 = \frac1q + 196884q + \ldots.
\ee
Holomorphic modular invariant functions with poles only at $\tau=i\infty$ are polynomials in $J(\tau)$.

Finally, we consider the subgroup of $SL(2,\bb{Z})$ called $\Gamma_0(2)$ defined as matrices $\bigl(\begin{smallmatrix}
a&b \\ c&d
\end{smallmatrix} \bigr) \in SL(2,\bb{Z})$ with $c$ even. Modular forms under $\Gamma_0(2)$ are generated by the functions $E_2'(\tau)$, defined as
\be
E_2'(\tau) = 1+24\sum_{n=1}^{\infty} \frac{nq^n}{1+q^n}
\ee
and $E_4(\tau)$, defined in (\ref{eq:e4blahblah}).

\section{Current Algebra and Flavored Partition Function}
\label{app:transformationderivation}

We have argued that the transformation property,
\es{trans}{
Z\left(\tau^{\prime},z^{\prime}\right)&=e^{\pi i \frac{c z^{2}}{c \tau+d}}Z(\tau,z)\,,
}
relies on the universal structure of the current algebra, rather than any theory specific details. Here we demonstrate this in gory detail.

\subsection{Perturbative Argument}
Our strategy will be to calculate the transformation property of the flavored partition function (\ref{trans}) order by order in $z$ about 0. The transformation rule can be verified at each order using the structure of the current algebra without any knowledge of the particular theory. We demonstrate this explicitly at quadratic order in $z$ and then present the general argument. As the rule is theory independent we can thus read it off from any theory we like, for instance the free boson, for which the rule (\ref{trans}) is well known (see \cite{Kraus:2006wn} for instance).
\\

\underline{Quadratic Order}:
\\

At quadratic order we have,
\es{dertrans}{
\partial_{z}^{2}Z(\tau^{\prime})\Big|_{z=0}&=(c \tau+d)^{2}\left(\partial_{z}^{2}Z(\tau)\Big|_{z=0}+2\pi i\frac{c}{c \tau+d}Z(\tau)\Big|_{z=0} \right)\,.
}
The $z$ derivatives are always evaluated at $z=0$, but we refrain from writing this below, to avoid clutter. We want to check this second order transformation by explicitly computing,
\es{Zpp1}{
\partial_{z}^{2}Z(\tau)&=(2\pi i)^{2}\tr\left(q^{L_{0}-c/24}J_{0}^{2}\right)\,.
}
In order to do this, we would like to find a primary that contains $J_{0}^{2}$ as part of its zero mode, as well as other known contributions. This is convenient as we know how primary one point functions transform, and thus can solve for the transformation of $\partial_{z}^{2}Z(\tau)$. Such an operator is given by,
\es{primop}{
\mathcal{O}_{2}(z)&=J^{2}(z)-\frac{2}{c}T(z) \ \ \ \leftrightarrow \ \ \ \left(J_{-1}^{2}-\frac{2}{c}L_{-2}\right)|0\rangle\,,
}
which has a zero mode,
\es{Jsqzm}{
(\mathcal{O}_{2})_{0}&=J_{0}^{2}+2\sum_{n\geq1}J_{-n}J_{n}-\frac{2}{c}L_{0}\,.
}
We can compute the torus one point function of $\mathcal{O}_{2}$\,.\footnote{This style of computation is similar to that presented in \cite{Iles:2013jha}, for example.}
\es{t1pt}{
F_{\mathcal{O}_{2}}(\tau)&\equiv(2\pi i)^{2}\tr\left(q^{L_{0}-c/24}(\mathcal{O}_{2})_{0}\right)\\
&=\underbrace{(2\pi i)^{2}\tr\left(q^{L_{0}-c/24}J^{2}_{0}\right)}_{\partial_{z}^{2}Z(\tau)}+2(2\pi i)^{2}\sum_{n\geq1}\tr\left(q^{L_{0}-c/24}J_{-n}J_{n}\right)-\frac{2}{c}(2\pi i)^{2}\tr\left(q^{L_{0}-c/24}L_{0}\right)\,.
}
The second and third terms on the second line can be simplified. Starting with the third term we have,
\es{dzterm}{
\tr\left(q^{L_{0}-c/24}L_{0}\right)&=q^{-c/24}\left(q\partial_{q}(q^{c/24}Z(\tau))\right)\\
&=\partial_{\tau}Z(\tau)+\frac{c}{24}Z(\tau)\,,
}
while for the second term we use,
\es{simpsum}{
\tr\left(q^{L_{0}-c/24}J_{-n}J_{n}\right)&=q^{n}\tr\left(q^{L_{0}-c/24}J_{n}J_{-n}\right)\\
&=\frac{nq^{n}}{1-q^{n}}Z(\tau)\,,
}
and the definition of the Eisenstein series to write,
\es{summedexp}{
\sum_{n\geq1}\tr\left(q^{L_{0}-c/24}J_{-n}J_{n}\right)&=\frac{1-E_{2}(\tau)}{24}Z(\tau)\,.
}
Putting this together, we can solve for $\partial_{z}^{2}Z(\tau)$.
\es{zpp2}{
\partial_{z}^{2}Z(\tau)&=F_{\mathcal{O}_{2}}(\tau)+(2\pi i)^{2}\left(\frac{E_{2}(\tau)}{12}+\frac{2}{c}\partial_{\tau}\right)Z(\tau)\,.
}
We are now in a position to write down the transformation properties of $\partial_{z}^{2}Z(\tau)$\,.
\es{quadtrans}{
\partial_{z}^{2}Z\left(\tau^{\prime}\right)&=(c \tau+d)^{2}\partial_{z}^{2}Z(\tau)+2\pi i c(c\tau+d)Z(\tau),
}
as desired.

In deriving this, we used the fact that both $F_{\mathcal{O}_{2}}(\tau)$ and $\partial_{\tau}Z(\tau)$ are modular forms of weight 2, as well as the anomalous transformation of $E_{2}(\tau)$ written in (\ref{eq:e2trans}).
\\

\underline{General Order}:
\\

To compute at arbitrary order we can replicate the argument style used above. To compute $\tr\left(q^{L_{0}-c/24}J_{0}^{m}\right)$, we look for a primary operator which contains $J_{0}^{m}$ as part of its zero mode. In addition it will contain terms of weight zero built out of $L_{m}$ and $J_{m}$ modes. The traces over these terms can be evaluated, as they were in the quadratic case, using only the current algebra to reduce them to modular differential operators acting on traces with fewer powers of $J_{0}$. Thus the modular properties of $\tr\left(q^{L_{0}-c/24}J_{0}^{m}\right)$ only depend on the universal current algebra, and so at each order, the transformation rule is identical in any theory. In particular, we can compute the transformation rule in the case of the free boson. This gives (\ref{trans}), and so it must also be correct for any theory with a $U(1)$ symmetry.

\section{Transformation of Characters}
\label{app:chartrans}
Modular invariance can be thought of as a sharp relation between the UV and the IR spectrum of the theory.  One way to build some additional intuition on the relation in a general theory is to look at the image under $S:\tau\rightarrow -\frac{1}{\tau}$ of an individual character.  In \cite{Keller:2014xba}, the transformation of characters of the Virasoro algebra were derived.  One might hope that further development of this approach to include the image under the full modular group could allow one to construct representations of Virasoro plus modular invariance, which could then be used as modules to be added to add additional states the full partition function.  In the case of chiral theories, Rademacher sums indeed make this a viable and useful method.  In the general non-holomorphic case, the major obstacle is that the image under $S$ produces a continuous, rather than a discrete, spectrum, and it is not clear how to systematically correct this.  Moreover, since the image of a single character is an integral over a continuum of characters up to arbitrarily high weight, for the analysis to be ``closed'' in a sense one must also characterize the modular image of infinite sums over characters as well.  Despite these caveats, we find the results of \cite{Keller:2014xba} to provide some useful guidance in thinking about modular transformations of non-holomorphic theories.   In this appendix, we will therefore consider the modular transformation of an individual Virasoro $\times$ current algebra character, which we describe below. We assume the existence of both left and right $U(1)$ currents for ease of exposition. 

The holomorphic Virasoro $\times$ $U(1)$ Affine Kac-Moody algebra is given by,
\be
\left[ L_m, L_n \right] &=& (m-n)L_{m+n} + \frac{c}{12} m(m^2-1) \delta_{m+n,0},  \nn\\
\left[ L_m, J_n \right] &=& -n J_{n+m} , \nn\\
\left[ J_m, J_n \right] &=& m k \delta_{n+m,0}\,,
\ee
and similarly for the anti-holomorphic algebra.
If $c>2$, the full irreducible representations of the Virasoro and current algebra are generated by all combinations of $J_{-n}, n\ge 1$ and $L_{-n}, n \ge 2$, acting on the primary states, as well as the anti-holomorphic modes.  Since these do not change the total $U(1)$ charge, and they raise the $L_0$ eigenvalue by $n$, one can immediately write the characters as products of the characters $\chi_J$ and $\chi_T$ under the two sectors separately:
\be
\chi_J(q) &=& \prod_{n=1}^\infty \frac{1}{1-q^n}, \nn\\
\chi_T(q)&=& q^{h}\left( \prod_{n=1}^\infty \frac{1}{1-q^n} \right) \left\{ \begin{array}{cc} 1-q & \textrm{vacuum} \\
1 & h>-\frac{c}{24} \end{array} \right\}
\ee
The full character is 
\be
\chi_{h,Q,c}(q,y) = y^Q \chi_T(q)\chi_J(q)  \bar{y}^{\bar Q} \bar{\chi}_T(\bar{q})\bar{\chi}_J(\bar{q})
\label{eq:fullchar}
\ee
where we have graded over the $U(1)$ left- and right-moving charges $Q,\bar{Q}$ with $y=e^{2\pi i z}$.  It is convenient to multiply by the modular invariant  function $\left| (i \tau)^{1/4} \eta(\tau)  \right|^4 $ to get the ``reduced'' characters:
\be
\hat{\chi}(q,y) &=&|\tau|(q \bar{q})^{\frac{1}{12}}  \left\{ \begin{array}{cc}  y^Q \bar{y}^{\bar{Q}} q^{h} \bar{q}^{\bar{h}}  & h>-\frac{c}{24}, \bar{h}>-\frac{\bar{c}}{24}, \\
(1-q)(1-\bar{q}) & \textrm{vacuum} \end{array} \right\}.
\ee

We want to consider what happens if we add an extra non-vacuum state to a theory.  We can focus on the left-moving part of the reduced character
\be
\hat\chi(\tau,z) &=& (i\tau)^{1/2} e^{ 2 \pi i(\tau E_L + z Q)}.
\ee 
Under $S$, this character gets mapped to
\be
e^{-2 \pi i \frac{c}{6} \frac{z^2}{\tau}} (i\tau)^{-1/2}e^{-2 \pi i E_L/\tau} e^{\frac{2\pi i z Q}{\tau}}.
\ee
Our goal is to decompose this into an integral over the untransformed characters times a density of states $\rho(E,Q)$:
\be
\int d E_L' dQ' \rho(E_L',Q') (i\tau)^{1/2} e^{2 \pi i (\tau E_L' + z Q')  }
\ee
Integrating both sides against $\int dz e^{-2 \pi i Q''}$, we obtain
\be
\int dE_L' e^{2 \pi i \tau E_L'} \rho(E', Q'') &=& \frac{1}{(i\tau)}e^{-2 \pi i E_L/\tau}  \int dz e^{-2 \pi i  \left( \frac{c}{6} \frac{z^2}{\tau}+  Q'' z -\frac{z Q}{\tau} \right)} \nn\\
 &=&  - \frac{\sqrt{3}}{\sqrt{-i c \tau}} e^{- 2 \pi i E_L/\tau} e^{\frac{3 \pi i (Q-Q'' \tau)^2}{c \tau}}
\ee 
This is just the left-moving piece of the full character; multiplying by the corresponding right-moving piece, we find
\begin{eqnarray}
\int dE_L' d E_R' \rho(E_L', E_R', Q'', \bar{Q}'') e^{2 \pi i (\tau E_L' + \bar{\tau} E_R')} =
\frac{3}{c |\tau|}  e^{-2 \pi i (\frac{E_L}{\tau} + \frac{E_R}{\bar{\tau}})} e^{\frac{3 \pi i (Q-Q'' \tau)^2}{c \tau}+\frac{3 \pi i (\bar{Q}-\bar{Q}'' \bar{\tau})^2}{c \bar{\tau}}}
\end{eqnarray}
If we assume that the theory satisfies charge conjugation symmetry, then for each state with charge $(Q, \bar{Q})$ and energy $(E_L, E_R)$, there is another state with charge $(-Q,-\bar{Q})$ and energy $(E_L, E_R)$.  Adding these two contributions together,  their image under $S$ has a spectrum given by
\be
\int dE_L' d E_R' \rho(E_L', E_R', Q'', \bar{Q}'') e^{2 \pi i (\tau E_L' + \bar{\tau} E_R')} &=&
\frac{3}{c |\tau|}  e^{-2 \pi i (\frac{E_L}{\tau} + \frac{E_R}{\bar{\tau}})} e^{\frac{3 \pi i (Q^2+Q''^2 \tau^2)^2}{c \tau} +\frac{3 \pi i (\bar{Q}^2+\bar{Q}''^2 \bar{\tau})^2}{c \bar{\tau}}} \nn\\
&& \times 2 \cos \left( \frac{ 6 \pi }{c} (Q Q'' + \bar{Q} \bar{Q}'')\right) 
\ee
To bring this into a more natural form, we can massage it a little to be
\be
\int dE_L' dE_R' \rho(E_L', E_R', Q', \bar{Q}')  e^{2 \pi i \left(\tau (E_L'-  \frac{3Q'^2}{2c}) + \bar{\tau} (E_R' -\frac{3 \bar{Q}'^2}{2c})\right) }  &=& \frac{3 }{c |\tau|} e^{- 2 \pi i \left( \frac{1}{\tau}(E_L - \frac{3 Q^2}{2c}) + \frac{1}{\bar{\tau} }(E_R - \frac{3 \bar{Q}^2}{2c} )\right)}\nn\\
& \times& 2 \cos \left( \frac{ 6 \pi }{c} (Q Q' + \bar{Q} \bar{Q}')\right) .
\ee
Clearly, it is natural to define the variables
\be
\tilde{E}_L  \equiv E_L - \frac{3Q^2}{2c}, \qquad \tilde{E_R} \equiv E_R - \frac{3\bar{Q}^2}{2c}.
\ee
In terms of these variables, the above relation takes the simple form
\be
\int d \tilde{E}_L' d \tilde{E}_R' \rho(\tilde{E}_L', \tilde{E}_R', Q', \bar{Q}')e^{2 \pi i ( \tau \tilde{E}_L' + \bar{\tau} \tilde{E}_R')}
=
\frac{6}{c |\tau|} e^{-2 \pi i \left( \frac{\tilde{E}_L}{\tau} + \frac{\tilde{E}_R}{\tau}\right) }  \cos \left( \frac{ 6 \pi  (Q Q' + \bar{Q} \bar{Q}')}{c}\right) . \nn\\
\ee
This has reduced to the transformation for the case $Q=0$, up to  an extra $\cos$ factor, and with the $E$'s are replaced by $\tilde{E}$'s.    But that is exactly the transformation that was derived in \cite{Keller:2014xba}\footnote{See their equation (23).}
Adopting their result (and keeping track of our slightly different integration measure), we finally arrive at
\be
\rho(\tilde{E}_L', \tilde{E}_R', Q', \bar{Q}') &=& \frac{12 }{c} \Theta(\tilde{E}_L') \Theta(\tilde{E}_R') \frac{1}{\sqrt{\tilde{E}_L' \tilde{E}_R'}}\cosh(4 \pi i \sqrt{ \tilde{E}_L \tilde{E}_L'}) \cosh( 4 \pi i \sqrt{\tilde{E}_R \tilde{E}_R'}) \nn\\
 && \times \cos \left( \frac{ 6 \pi }{c} (Q Q' + \bar{Q} \bar{Q}')\right).
\ee

\end{appendices}

\newpage

\bibliographystyle{utphys}
\bibliography{refs}

\begin{thebibliography}{10}
\ifx\href\asklfhas\newcommand{\href}[2]{#2}\fi
\ifx\arxivref\asklfhas\newcommand{\arxivref}[2]{\href{http://arxiv.org/abs/#1}{#2}}\fi
\ifx\doiref\asklfhas\newcommand{\doiref}[2]{\href{http://dx.doi.org/#1}{#2}}\fi
\parskip 0pt
\normalsize

\bibitem{Adams:2006sv}
A.~Adams, N.~Arkani-Hamed, S.~Dubovsky, A.~Nicolis \& R.~Rattazzi,
\textit{``{Causality, analyticity and an IR obstruction to UV completion}''},
\doiref{10.1088/1126-6708/2006/10/014}{JHEP \textbf{0610}, 014 (2006)},
\normalsize{\texttt{\arxivref{hep-th/0602178}{hep-th/0602178}}}.

\bibitem{GKP}
S.~S. Gubser, I.~R. Klebanov \& A.~M. Polyakov,
\textit{``{Gauge theory correlators from non-critical string theory}''},
\doiref{10.1016/S0370-2693(98)00377-3}{Phys.~Lett. \textbf{B428}, 105 (1998)},
\normalsize{\texttt{\arxivref{hep-th/9802109}{hep-th/9802109}}}.

\bibitem{Witten}
E.~Witten,
\textit{``{Anti-de Sitter space and holography}''},
Adv.~Theor.~Math.~Phys. \textbf{2}, 253 (1998),
\normalsize{\texttt{\arxivref{hep-th/9802150}{hep-th/9802150}}}.

\bibitem{Maldacena:1997re}
J.~M. Maldacena,
\textit{``{The large N limit of superconformal field theories and
  supergravity}''},
Adv.~Theor.~Math.~Phys. \textbf{2}, 231 (1998),
\normalsize{\texttt{\arxivref{hep-th/9711200}{hep-th/9711200}}}.

\bibitem{Fitzpatrick:2014vua}
A.~L. Fitzpatrick, J.~Kaplan \& M.~T. Walters,
\textit{``{Universality of Long-Distance AdS Physics from the CFT
  Bootstrap}''},
\doiref{10.1007/JHEP08(2014)145}{JHEP \textbf{1408}, 145 (2014)},
\normalsize{\texttt{\arxivref{1403.6829}{arXiv:1403.6829}}}.

\bibitem{Fitzpatrick:2010zm}
A.~L. Fitzpatrick, E.~Katz, D.~Poland \& D.~Simmons-Duffin,
\textit{``{Effective Conformal Theory and the Flat-Space Limit of AdS}''},
\doiref{10.1007/JHEP07(2011)023}{JHEP \textbf{1107}, 023 (2011)},
\normalsize{\texttt{\arxivref{1007.2412}{arXiv:1007.2412}}}.

\bibitem{Komargodski:2012ek}
Z.~Komargodski \& A.~Zhiboedov,
\textit{``{Convexity and Liberation at Large Spin}''},
\doiref{10.1007/JHEP11(2013)140}{JHEP \textbf{1311}, 140 (2013)},
\normalsize{\texttt{\arxivref{1212.4103}{arXiv:1212.4103}}}.

\bibitem{Fitzpatrick:2012yx}
A.~L. Fitzpatrick, J.~Kaplan, D.~Poland \& D.~Simmons-Duffin,
\textit{``{The Analytic Bootstrap and AdS Superhorizon Locality}''},
\doiref{10.1007/JHEP12(2013)004}{JHEP \textbf{1312}, 004 (2013)},
\normalsize{\texttt{\arxivref{1212.3616}{arXiv:1212.3616}}}.

\bibitem{Alday:2007mf}
L.~F. Alday \& J.~M. Maldacena,
\textit{``{Comments on operators with large spin}''},
\doiref{10.1088/1126-6708/2007/11/019}{JHEP \textbf{0711}, 019 (2007)},
\normalsize{\texttt{\arxivref{0708.0672}{arXiv:0708.0672}}}.

\bibitem{Hellerman}
S.~Hellerman,
\textit{``{A Universal Inequality for CFT and Quantum Gravity}''},
\doiref{10.1007/JHEP08(2011)130}{JHEP \textbf{1108}, 130 (2011)},
\normalsize{\texttt{\arxivref{0902.2790}{arXiv:0902.2790}}}.

\bibitem{Hartman:2014oaa}
T.~Hartman, C.~A. Keller \& B.~Stoica,
\textit{``{Universal Spectrum of 2d Conformal Field Theory in the Large c
  Limit}''},
\doiref{10.1007/JHEP09(2014)118}{JHEP \textbf{1409}, 118 (2014)},
\normalsize{\texttt{\arxivref{1405.5137}{arXiv:1405.5137}}}.

\bibitem{Benjamin:2015vkc}
N.~Benjamin, S.~Kachru, C.~A. Keller \& N.~M. Paquette,
\textit{``{Emergent space-time and the supersymmetric index}''},
\normalsize{\texttt{\arxivref{1512.00010}{arXiv:1512.00010}}}.

\bibitem{Benjamin:2015hsa}
N.~Benjamin, M.~C.~N. Cheng, S.~Kachru, G.~W. Moore \& N.~M. Paquette,
\textit{``{Elliptic Genera and 3d Gravity}''},
\normalsize{\texttt{\arxivref{1503.04800}{arXiv:1503.04800}}}.

\bibitem{Friedan:2013cba}
D.~Friedan \& C.~A. Keller,
\textit{``{Constraints on 2d CFT partition functions}''},
\doiref{10.1007/JHEP10(2013)180}{JHEP \textbf{1310}, 180 (2013)},
\normalsize{\texttt{\arxivref{1307.6562}{arXiv:1307.6562}}}.

\bibitem{Qualls:2015bta}
J.~D. Qualls,
\textit{``{Universal Bounds on Operator Dimensions in General 2D Conformal
  Field Theories}''},
\normalsize{\texttt{\arxivref{1508.00548}{arXiv:1508.00548}}}.

\bibitem{Qualls:2014oea}
J.~D. Qualls,
\textit{``{Universal Bounds in Even-Spin CFTs}''},
\doiref{10.1007/JHEP12(2015)001}{JHEP \textbf{1512}, 001 (2015)},
\normalsize{\texttt{\arxivref{1412.0383}{arXiv:1412.0383}}}.

\bibitem{Qualls:2013eha}
J.~D. Qualls \& A.~D. Shapere,
\textit{``{Bounds on Operator Dimensions in 2D Conformal Field Theories}''},
\doiref{10.1007/JHEP05(2014)091}{JHEP \textbf{1405}, 091 (2014)},
\normalsize{\texttt{\arxivref{1312.0038}{arXiv:1312.0038}}}.

\bibitem{BTZ}
M.~Banados, C.~Teitelboim \& J.~Zanelli,
\textit{``{The Black hole in three-dimensional space-time}''},
\doiref{10.1103/PhysRevLett.69.1849}{Phys.~Rev.~Lett. \textbf{69}, 1849
  (1992)},
\normalsize{\texttt{\arxivref{hep-th/9204099}{hep-th/9204099}}}.

\bibitem{ArkaniHamed:2006dz}
N.~Arkani-Hamed, L.~Motl, A.~Nicolis \& C.~Vafa,
\textit{``{The String landscape, black holes and gravity as the weakest
  force}''},
\doiref{10.1088/1126-6708/2007/06/060}{JHEP \textbf{0706}, 060 (2007)},
\normalsize{\texttt{\arxivref{hep-th/0601001}{hep-th/0601001}}}.

\bibitem{Banks:2010zn}
T.~Banks \& N.~Seiberg,
\textit{``{Symmetries and Strings in Field Theory and Gravity}''},
\doiref{10.1103/PhysRevD.83.084019}{Phys.~Rev. \textbf{D83}, 084019 (2011)},
\normalsize{\texttt{\arxivref{1011.5120}{arXiv:1011.5120}}}.

\bibitem{Harlow:2015lma}
D.~Harlow,
\textit{``{Wormholes, Emergent Gauge Fields, and the Weak Gravity
  Conjecture}''},
\doiref{10.1007/JHEP01(2016)122}{JHEP \textbf{1601}, 122 (2016)},
\normalsize{\texttt{\arxivref{1510.07911}{arXiv:1510.07911}}}.

\bibitem{Heidenreich:2015nta}
B.~Heidenreich, M.~Reece \& T.~Rudelius,
\textit{``{Sharpening the Weak Gravity Conjecture with Dimensional
  Reduction}''},
\doiref{10.1007/JHEP02(2016)140}{JHEP \textbf{1602}, 140 (2016)},
\normalsize{\texttt{\arxivref{1509.06374}{arXiv:1509.06374}}}.

\bibitem{Nakayama:2015hga}
Y.~Nakayama \& Y.~Nomura,
\textit{``{Weak gravity conjecture in the AdS/CFT correspondence}''},
\doiref{10.1103/PhysRevD.92.126006}{Phys.~Rev. \textbf{D92}, 126006 (2015)},
\normalsize{\texttt{\arxivref{1509.01647}{arXiv:1509.01647}}}.

\bibitem{Brown:1986nw}
J.~D. Brown \& M.~Henneaux,
\textit{``{Central Charges in the Canonical Realization of Asymptotic
  Symmetries: An Example from Three-Dimensional Gravity}''},
\doiref{10.1007/BF01211590}{Commun.~Math.~Phys. \textbf{104}, 207 (1986)}.

\bibitem{Cheung:2014ega}
C.~Cheung \& G.~N. Remmen,
\textit{``{Infrared Consistency and the Weak Gravity Conjecture}''},
\doiref{10.1007/JHEP12(2014)087}{JHEP \textbf{1412}, 087 (2014)},
\normalsize{\texttt{\arxivref{1407.7865}{arXiv:1407.7865}}}.

\bibitem{Dijkgraaf:1987vp}
R.~Dijkgraaf, E.~P. Verlinde \& H.~L. Verlinde,
\textit{``{C = 1 Conformal Field Theories on Riemann Surfaces}''},
\doiref{10.1007/BF01224132}{Commun.~Math.~Phys. \textbf{115}, 649 (1988)}.

\bibitem{Verlinde:1986kw}
E.~P. Verlinde \& H.~L. Verlinde,
\textit{``{Chiral Bosonization, Determinants and the String Partition
  Function}''},
\doiref{10.1016/0550-3213(87)90219-7}{Nucl.~Phys. \textbf{B288}, 357 (1987)}.

\bibitem{Kiritsis:2007zza}
E.~Kiritsis,
\textit{``{String theory in a nutshell}''},
Princeton University Press (2007).

\bibitem{AlvarezGaume:1986es}
L.~Alvarez-Gaume, G.~W. Moore \& C.~Vafa,
\textit{``{Theta Functions, Modular Invariance and Strings}''},
\doiref{10.1007/BF01210925}{Commun.~Math.~Phys. \textbf{106}, 1 (1986)}.

\bibitem{Kawai:1993jk}
T.~Kawai, Y.~Yamada \& S.-K. Yang,
\textit{``{Elliptic genera and N=2 superconformal field theory}''},
\doiref{10.1016/0550-3213(94)90428-6}{Nucl.~Phys. \textbf{B414}, 191 (1994)},
\normalsize{\texttt{\arxivref{hep-th/9306096}{hep-th/9306096}}}.

\bibitem{RASOF1962165}
B.~Rasof,
\textit{``The initial- and final-value theorems in Laplace transform theory''},
\doiref{http://dx.doi.org/10.1016/0016-0032(62)90939-0}{Journal~of~the~Franklin~Institute
  \textbf{274}, 165  (1962)},
\href{http://www.sciencedirect.com/science/article/pii/0016003262909390}{\texttt{http://www.sciencedirect.com/science/article/pii/0016003262909390}}.

\bibitem{ModularStein}
W.~A. Stein,
\textit{``Modular Forms, A Computational Approach''},
American Mathematical Society (2007).

\bibitem{JenkinsRouse}
P.~Jenkins \& J.~Rouse,
\textit{``{Bounds for Coefficients of Cusp Forms and Extremal Lattices}''},
\normalsize{\texttt{\arxivref{1012.5991}{arXiv:1012.5991}}}.

\bibitem{Leech:1967}
J.~Leech,
\textit{``Notes on Sphere Packings''},
\doiref{http://dx.doi.org/10.4153/CJM-1967-017-0}{Canad.~J.~Math. \textbf{19},
  251  (1967)},
\href{http://cms.math.ca/10.4153/CJM-1967-017-0}{\texttt{http://cms.math.ca/10.4153/CJM-1967-017-0}}.

\bibitem{Nebe1998472}
G.~Nebe,
\textit{``Some Cyclo-quaternionic Lattices''},
\doiref{http://dx.doi.org/10.1006/jabr.1997.7163}{Journal~of~Algebra
  \textbf{199}, 472  (1998)},
\href{http://www.sciencedirect.com/science/article/pii/S0021869397971635}{\texttt{http://www.sciencedirect.com/science/article/pii/S0021869397971635}}.

\bibitem{Nebe:2010}
G.~Nebe,
\textit{``{An even unimodular 72-dimensional lattice of minimum 8}''},
J.~Reine~und~Angew.~Math. \textbf{673}, 237 (2012),
\normalsize{\texttt{\arxivref{1008.2862}{arXiv:1008.2862}}}.

\bibitem{Caselle:1988jm}
M.~Caselle \& K.~S. Narain,
\textit{``{A New Approach to the Construction of Conformal Field Theories}''},
\doiref{10.1016/0550-3213(89)90129-6}{Nucl.~Phys. \textbf{B323}, 673 (1989)}.

\bibitem{Dolan:1989vr}
L.~Dolan, P.~Goddard \& P.~Montague,
\textit{``{Conformal Field Theory of Twisted Vertex Operators}''},
\doiref{10.1016/0550-3213(90)90644-S}{Nucl.~Phys. \textbf{B338}, 529 (1990)}.

\bibitem{amwip}
N.~Benjamin, E.~Dyer, A.~L. Fitzpatrick, A.~Maloney \& E.~Perlmutter,
\textit{``{To appear}''}.

\bibitem{Rattazzi:2008pe}
R.~Rattazzi, V.~S. Rychkov, E.~Tonni \& A.~Vichi,
\textit{``{Bounding scalar operator dimensions in 4D CFT}''},
\doiref{10.1088/1126-6708/2008/12/031}{JHEP \textbf{0812}, 031 (2008)},
\normalsize{\texttt{\arxivref{0807.0004}{arXiv:0807.0004}}}.

\bibitem{Poland:2011ey}
D.~Poland, D.~Simmons-Duffin \& A.~Vichi,
\textit{``{Carving Out the Space of 4D CFTs}''},
\doiref{10.1007/JHEP05(2012)110}{JHEP \textbf{1205}, 110 (2012)},
\normalsize{\texttt{\arxivref{1109.5176}{arXiv:1109.5176}}}.

\bibitem{ElShowk:2012ht}
S.~El-Showk, M.~F. Paulos, D.~Poland, S.~Rychkov, D.~Simmons-Duffin \&
  A.~Vichi,
\textit{``{Solving the 3D Ising Model with the Conformal Bootstrap}''},
\doiref{10.1103/PhysRevD.86.025022}{Phys.~Rev. \textbf{D86}, 025022 (2012)},
\normalsize{\texttt{\arxivref{1203.6064}{arXiv:1203.6064}}}.

\bibitem{ElShowk:2012hu}
S.~El-Showk \& M.~F. Paulos,
\textit{``{Bootstrapping Conformal Field Theories with the Extremal Functional
  Method}''},
\doiref{10.1103/PhysRevLett.111.241601}{Phys.~Rev.~Lett. \textbf{111}, 241601
  (2013)},
\normalsize{\texttt{\arxivref{1211.2810}{arXiv:1211.2810}}}.

\bibitem{Kos:2014bka}
F.~Kos, D.~Poland \& D.~Simmons-Duffin,
\textit{``{Bootstrapping Mixed Correlators in the 3D Ising Model}''},
\doiref{10.1007/JHEP11(2014)109}{JHEP \textbf{1411}, 109 (2014)},
\normalsize{\texttt{\arxivref{1406.4858}{arXiv:1406.4858}}}.

\bibitem{Witten:2007kt}
E.~Witten,
\textit{``{Three-Dimensional Gravity Revisited}''},
\normalsize{\texttt{\arxivref{0706.3359}{arXiv:0706.3359}}}.

\bibitem{MooreLectures}
G.~Moore,
\textit{``{Trieste Lectures on Mathematical Aspects of Supersymmetric Black
  Holes}''},
\href{http://www.physics.rutgers.edu/~gmoore/TriesteLectures\_March28\_2008.pdf}{\texttt{http://www.physics.rutgers.edu/~gmoore/TriesteLectures\_March28\_2008.pdf}}.

\bibitem{Aharony:2013hda}
O.~Aharony, N.~Seiberg \& Y.~Tachikawa,
\textit{``{Reading between the lines of four-dimensional gauge theories}''},
\doiref{10.1007/JHEP08(2013)115}{JHEP \textbf{1308}, 115 (2013)},
\normalsize{\texttt{\arxivref{1305.0318}{arXiv:1305.0318}}}.

\bibitem{Keller:2012mr}
C.~A. Keller \& H.~Ooguri,
\textit{``{Modular Constraints on Calabi-Yau Compactifications}''},
\doiref{10.1007/s00220-013-1797-8}{Commun.~Math.~Phys. \textbf{324}, 107
  (2013)},
\normalsize{\texttt{\arxivref{1209.4649}{arXiv:1209.4649}}}.

\bibitem{Fiset:2015pta}
M.-A. Fiset \& J.~Walcher,
\textit{``{Bounding the Heat Trace of a Calabi-Yau Manifold}''},
\doiref{10.1007/JHEP09(2015)124}{JHEP \textbf{1509}, 124 (2015)},
\normalsize{\texttt{\arxivref{1506.08407}{arXiv:1506.08407}}}.

\bibitem{Kraus:2006wn}
P.~Kraus,
\textit{``{Lectures on black holes and the AdS(3) / CFT(2) correspondence}''},
Lect.~Notes~Phys. \textbf{755}, 193 (2008),
\normalsize{\texttt{\arxivref{hep-th/0609074}{hep-th/0609074}}}.

\bibitem{Iles:2013jha}
N.~J. Iles \& G.~M.~T. Watts,
\textit{``{Characters of the $W_3$ algebra}''},
\doiref{10.1007/JHEP02(2014)009}{JHEP \textbf{1402}, 009 (2014)},
\normalsize{\texttt{\arxivref{1307.3771}{arXiv:1307.3771}}}.

\bibitem{Keller:2014xba}
C.~A. Keller \& A.~Maloney,
\textit{``{Poincare Series, 3D Gravity and CFT Spectroscopy}''},
\doiref{10.1007/JHEP02(2015)080}{JHEP \textbf{1502}, 080 (2015)},
\normalsize{\texttt{\arxivref{1407.6008}{arXiv:1407.6008}}}.

\end{thebibliography}

\end{document}